\documentclass[11pt]{article}
\pdfoutput=1
\usepackage{myoptions}
\usepackage{jheppub}
\usepackage{float, extarrows, tikz-cd}
\usepackage{graphicx}
\usepackage{slashed}
\usepackage{tabularx,ragged2e}
\usepackage{amssymb}
\usepackage{amsmath,amssymb}
\usepackage{slashed}
\usepackage{hyperref}
\usepackage{caption}
\usepackage{xcolor}
\usepackage{dsfont}
\usepackage{verbatim}
\usepackage{subfig}
\usepackage{mathtools,xcolor,ytableau,amsfonts,tikz}
\usepackage{graphicx}
\usepackage{physics}
\usepackage{amsthm}
\usepackage{scalerel}

\newcolumntype{C}{>{\Centering\arraybackslash}X}

\usepackage{tikz}
\newcommand*{\boxcolor}{black}
\makeatletter
\renewcommand{\boxed}[1]{\textcolor{\boxcolor}{%
\tikz[baseline={([yshift=-1ex]current bounding box.center)}] \node [rectangle, minimum width=1ex,rounded corners,draw] {\normalcolor\m@th$\;\,\displaystyle#1\;\,$};}}
 \makeatother

\newcommand{\beq}{\begin{equation}\begin{aligned}}
\newcommand{\eeq}{\end{aligned}\end{equation}}

\newcommand{\twopt}{{\cal G}}

\newcommand{\bpm}{\begin{pmatrix}}
\newcommand{\epm}{\end{pmatrix}}

\newcommand{\del}{\delta}
\newcommand{\ex}[1]{\left\langle{#1}\right\rangle}

\renewcommand{\title}[1]{\vbox{\center\LARGE{#1}}\vspace{5mm}}
\renewcommand{\author}[1]{\vbox{\center#1}\vspace{5mm}}
\newcommand{\address}[1]{\vbox{\center\em#1}}

\ytableausetup{centertableaux,boxsize=.5em}

\begin{document}

\thispagestyle{empty}

\title{Effective description of sub-maximal chaos:\\ stringy effects for SYK scrambling}

\bigskip

\begin{center}

\author{Changha Choi,$^1$ Felix M.\ Haehl,$^2$ M\'{a}rk Mezei,$^3$ and G\'{a}bor S\'{a}rosi$^4$}

\address{1) Perimeter Institute for Theoretical Physics,
Waterloo, Ontario, N2L 2Y5, Canada}

\address{2) School of Mathematical Sciences, University of Southampton, SO17 1BJ, U.K.}

\address{3) Mathematical Institute, University of Oxford, Woodstock Road, Oxford, OX2 6GG, U.K.}

\address{4) CERN, Theoretical Physics Department, 1211 Geneva 23, Switzerland}

\vspace{0.5in}
    
{\tt cchoi@perimeterinstitute.ca, f.m.haehl@soton.ac.uk, mezei@maths.ox.ac.uk, gabor.sarosi@cern.ch}

\bigskip

\vspace{1cm}

\end{center}

\begin{abstract}
\noindent  It has been proposed that the exponential decay and subsequent power law saturation of out-of-time-order correlation functions can be universally described by collective `scramblon' modes. We develop this idea from a path integral perspective in several examples, thereby establishing a general formalism. After reformulating previous work on the Schwarzian theory and identity conformal blocks in two-dimensional CFTs relevant for systems in the infinite coupling limit with maximal quantum Lyapunov exponent, we focus on theories with sub-maximal chaos: we study the large-$q$ limit of the SYK quantum dot and chain, both of which are amenable to analytical treatment at finite coupling. In both cases we identify the relevant scramblon modes, derive their effective action, and find bilocal vertex functions, thus constructing an effective description of chaos. 
The final results can be matched in detail to stringy corrections to the gravitational eikonal S-matrix in holographic CFTs, including a stringy Regge trajectory, bulk to boundary propagators, and  multi-string effects that are unexplored holographically.
\end{abstract}

\newpage

\newpage

\setcounter{tocdepth}{3}
{}
\vfill
\tableofcontents

\newpage

\section{Introduction} 

One measure of chaos in quantum many body systems is the time dependence of out-of-time-ordered correlation functions (OTOCs). These have been proposed as measures of the quantum butterfly effect \cite{Shenker:2013pqa,Roberts:2014isa,Maldacena:2015waa}, and have been computed in many different contexts. Some of the most insightful calculations were performed in Sachdev-Ye-Kitaev quantum mechanics \cite{KitaevTalks,Sachdev:2015efa,Maldacena:2016hyu}, which exhibits maximal chaos in the strong coupling limit, but sub-maximal chaos away from this limit. The corrections to maximal chaos are of paramount interest in holography, as they correspond to stringy corrections to `eikonal' gravitational scattering \cite{Shenker:2014cwa}. In order to see the emergence of a string theory description from boundary quantum dynamics, it is thus interesting to study the mechanism that leads to sub-maximal chaos in a language that is adapted to holography. 

To study these problems, our goal will be to formulate an effective description of quantum chaos in terms of variables that have a gravitational or stringy analog. Attempts to formulate an effective theory of chaos are not new: the paradigm for this idea is the Schwarzian theory, describing the strong coupling dynamics of the SYK model (and other models with similar symmetry breaking pattern) \cite{Maldacena:2016hyu,Jensen:2016pah} as well as the boundary degree of freedom of JT gravity \cite{Maldacena:2016upp}: it is an effective field theory for the time reparametrization mode whose effects are enhanced at strong coupling. Since the reparametrization mode plays an important role in nearly-AdS$_2$ gravity, formulating quantum chaos in terms of its dynamics makes the emergence of a gravitational description manifest. Similar discussions have appeared for other maximally chaotic systems such as two-dimensional CFTs, where boundary reparametrization modes similarly acquire an action that describe the `gravitational' contribution to OTOCs \cite{Turiaci:2016cvo,Cotler:2018zff,Haehl:2018izb}; see also \cite{Blake:2017ris,Blake:2021wqj} for a general effective theory for maximal chaos.
However, away from the maximal chaos limit, the same ideas using local effective field theory methods naively fail (e.g.\ \cite{Blake:2021wqj}). Recently a proposal was made for an effective description of scrambling that sidesteps the use of a local effective action: using path integrals for simple models \cite{Stanford:2021bhl} as well as Feynman diagrams in SYK-like systems \cite{Gu:2021xaj} a description in terms of 
scramblon amplitudes was proposed. Our goal will be to generalize these works and establish a technique, which should be applicable whenever a description of the system in terms of path integrals (or an effective action) is available. 

For concreteness we will focus on specific models, which nevertheless will display a high degree of generality. Our main attention will be on the large $q$ limit of the SYK model: after taking the usual large $N$ limit, we consider an expansion in $\frac{1}{q}$, where $q$ denotes the number of fermions interacting through any realization of random couplings (i.e., we take $N \gg q \gg 1$).\footnote{ This is distinct from the double scaling limit, where $\frac{q^2}{N}$ is held fixed \cite{Berkooz:2018qkz,Berkooz:2018jqr}.} In the large $q$ limit many of the finite coupling effects become exactly solvable, allowing us to derive the effective description of chaos in detail. For instance, the OTOC at order $1/N$  is known exactly \cite{Streicher:2019wek,Choi:2019bmd}, and it exhibits exponential growth with Lyapunov exponent $\frac{2\pi v}{\beta}$ where $0< v \leq 1$ parametrizes the coupling strength. Correlation functions in the large $q$ model display an `extended range of analyticity' due to the effective rescaling of inverse temperature $\beta \rightarrow \frac{\beta}{v}$. Despite this peculiarity, we shall nevertheless give arguments that the deviation from maximal chaos is indeed analogous to stringy corrections to the gravitational eikonal S-matrix. This will be particularly clear when generalizing to a nearest-neighbor chain of large $q$ SYK quantum `dots', as the spatial dependence plays a crucial role in the comparison with stringy effects in gravity \cite{Choi:2020tdj}.

In CFTs with a sting theory dual with small but finite string length the OTOC takes the following schematic form \cite{Shenker:2014cwa}:
\es{StringyEffects}{
\text{OTOC}(x_i)=\int \prod_{i=1}^4\psi_i(p_i,x_i) \exp\le[-iG_N\int dp \ {s^{J(p)-1}\ov D(p)} e^{i p b}\ri]\,,
}
where the $x_i$ are four spacetime positions of the external operators,  $\psi_i$ are bulk to boundary wave functions, $s$ is the Mandelstam invariant controlling scattering, $J(p)$ is the string Regge trajectory parametrized by the transverse momentum $p$, and $b$ is the impact parameter. Our final result in the SYK chain~\eqref{eq:FeikonalGrav2} is strikingly similar to this formula as we explain there; the spatial structure of the bulk to boundary wave functions is simpler and we encounter multi-string effects that are only suppressed close to the maximal chaos regime and are so far unexplored in holography. 

{\it Note added:} While we were finalizing this draft, we received~\cite{Gao:2023wun} that proposes 
an effective field theory for non-maximal quantum chaos, and which appears in the same {arXiv} posting. We did not have time to compare our approaches, but superficially they look different.

\section{Strategy: OTOCs in the eikonal limit from path integrals}

\begin{figure}
       \begin{center}
       \includegraphics[width=0.65\linewidth]{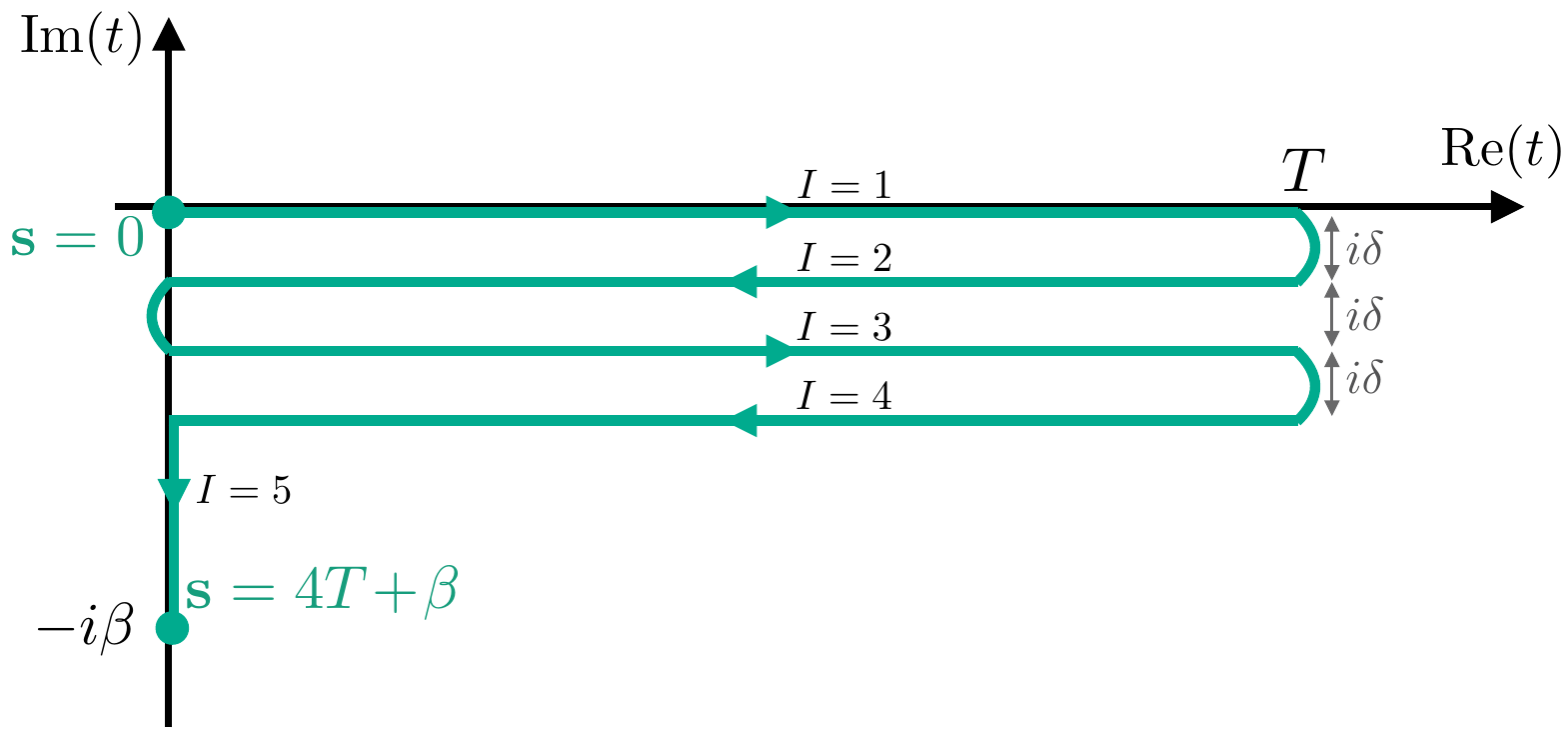}
       \end{center}
     \caption{Generalized Schwinger-Keldysh contour ${\cal C}$ for the path integral calculation of the out-of-time-order correlators. The integration contour in the complex $t$ plane is parametrized by the real-valued contour time $s \in [0,4T+\beta]$. Infinitesimal contour separations by $i\delta$ serve as a regulator (which we only keep track of when it is needed for regularization).}
     \label{fig:SK}
\end{figure}

Our goal is to present a novel way of computing out-of-time-order correlators in systems with sub-maximal chaos. We are hence interested in the following four-point function:
 \es{OTOCdef}{
 \langle T_{\cal C} \{ \sO_1(t_1)\sO_2(t_2) \sO_3(t_3) \sO_4(t_4) \}\rangle_\beta
 }
where $t_i$ are complex insertion times, $T_{\cal C}$ denotes time ordering along the generalized Schwinger-Keldysh contour ${\cal C}$ (see figure \ref{fig:SK}).\footnote{The particular SK contour we choose here is out of convenience, our methods straightforwardly generalize to SK contours with more Euclidean section, like the one appropriate for the regularized OTOC introduced in~\cite{Maldacena:2015waa}. A more symmetric contour configuration has the advantage that the OTOC is real. We find our choice more convenient for some calculations, but as a consequence the eikonal action and the OTOC are complex, typically by means of phase factors $e^{-\frac{i\pi v}{2}}$.} In SYK-type models the most elementary such correlator is the four-point functions of fermions.
We will primarily be interested in the out-of-time-order configuration, which corresponds to insertion times $\hat{t}_i$ with alternating insertions along the contour ${\cal C}$ (see figure \ref{fig:shocks}). Using a convenient normalization, we define it as
\begin{equation}
\label{eq:OTOCconfigDef}
\begin{split}
& \text{OTOC}(\hat{t}_1,\hat{t}_2,\hat{t}_3,\hat{t}_4) \equiv    \frac{1}{N^2} \sum_{i,j=1}^N \frac{ \langle T_{\cal C} \{ \psi_i(\hat{t}_1)\psi_i(\hat{t}_2) \psi_j(\hat{t}_3) \psi_j(\hat{t}_4) \}\rangle_\beta }{\twopt(\hat{t}_{12}) \twopt(\hat{t}_{34})} = 1 + \frac{1}{N} \, {\cal F}^\text{otoc} + {\cal O} \left( \frac{1}{N^2} \right)
\end{split}
\end{equation}
where $\twopt(t)=\langle T_{\cal C}\{\psi_i(t)\psi_i(0)\} \rangle$ denotes the two-point function. In the literature it is ${\cal F}^\text{otoc}$ that is analyzed most extensively;\footnote{Note a peculiarity of our conventions: we define ${\cal F}^\text{otoc}$ as the ${\cal O}(\frac{1}{N})$ piece {\it after} factoring out two-point functions that satisfy $\twopt(0)=\frac{1}{2}$.} in this paper we go beyond this to study the full OTOC.

We will specialize to the configuration $\hat{t}_1 = \hat{t}_2 = 0 \,,\; \hat{t}_3 = \hat{t}_4 = T$ to reduce clutter.\footnote{ We shall always denote the {\it specific} insertion points $\hat{t}_1 = \hat{t}_2 = 0 \,,\; \hat{t}_3 = \hat{t}_4 = T$ with a hat ($\,\hat{\;\;}\,$) to distinguish them from generic time arguments. When necessary, we regularize these insertion times using small imaginary parts $-i\delta$ as shown in figures \ref{fig:SK} and \ref{fig:shocks}. This will only be necessary in theories with maximal chaos.} 
To keep track of this, it will be useful to introduce a real-valued contour time $s \in [0,4T+\beta]$, which increases along the contour such that
\begin{equation}
 t_I(s) = \left\{ \begin{aligned} 
  \!\!s\,\, \qquad\qquad\qquad &(0\leq s \leq T) \\
  2T-s \qquad\qquad& ( T \leq s \leq 2T) \\
  s-2T \qquad\qquad& (2T \leq s \leq 3T) \\
  4T-s \qquad\qquad& (3T \leq s \leq 4T) \\
  -i(s-4T) \qquad& (4T \leq s \leq 4T+ \beta)
  \end{aligned}
  \right.
\end{equation}
The OTOC corresponds to the configuration 
\begin{equation}
\label{eq:OTOCconfig}
\begin{split}
& \text{OTOC:}  \qquad \hat{s}_4 =T \,,\quad \hat{s}_3 = 3T \,,\quad \hat{s}_2 = 2T \,,\quad \hat{s}_1 =4T \,.
\end{split}
\end{equation}

Our goal is to evaluate the OTOC in the eikonal approximation. This means we will be concerned with the leading contributions at large $T$. In particular, we will resum contributions of order $\big(\frac{1}{N} \, e^{\kappa T}\big)^n$, where $\kappa$ is the Lyapunov exponent. Our starting point is the path integral representation of the OTOC in terms of its large $N$ effective action on the generalized Schwinger-Keldysh contour:
\begin{equation}
  \text{OTOC}(t_1,t_2,t_3,t_4) = \int {\cal D}\phi \; e^{iS[\phi]} \; {\cal B}_\phi(t_1,t_2) \, {\cal B}_\phi(t_3,t_4)
\end{equation}
where ${\cal B}$ are bilocal vertices. 
In the SYK context, in the simplest case
${\cal B}=G(t_1,t_2)$ is the fermion bilinear, and $\phi$ represents the $G-\Sigma$ path integral.

As we will see, for times $T \sim \log N$, the dominant contributions to the path integral originate from a particular set of nearly-zero modes of the action: after constructing the saddle point solution, we will consider the quadratic action of fluctuations. The latter has a set of modes, which are almost zero modes if it were not for boundary contributions arising from discontinuities in the field configurations. We will argue that these dominate the OTOC computation, and we will reduce the computation of the OTOC to an integral over scalar parameters $X^\pm$ parametrizing the nearly-zero modes: 
\begin{equation}
\label{eq:OTOCeikonal}
\text{OTOC}  \approx  \text{OTOC}_{eik} = \int dX^+ dX^- \; e^{iS_{eik}[X^+,X^-]} \; {\cal B}_+(t_1,t_2) \, {\cal B}_-(t_3,t_4)
\end{equation}
It is useful to think of $X^\pm$ as quantifying the strength of `shocks' sourced by the operator insertions. The bilocal vertex functions ${\cal B}_\pm(t,t') \equiv {\cal B}_{X^\pm}(t,t')$ correspond to the two-point functions in the presence of the sources $X^\pm$. In the remainder of the paper we derive both the eikonal action, as well as the vertex functions $g_\pm$ for increasingly complicated systems, and we will verify that we reproduce the long time behavior of the OTOC in the cases where it has been computed before, as well as produce new results.

\section{Warmup: systems with maximal Lyapunov exponent}
\label{sec:maximal}

For maximally chaotic systems, the eikonal action and its use in the calculation for the OTOC has been understood for some time. In this section we shall review these developments in a slightly new language, which will set the stage for our later analysis of systems with sub-maximal chaos. We begin with the Schwarzian theory, for which these considerations were first discussed in \cite{Maldacena:2016upp}. We then discuss the case of the identity conformal block in two-dimensional CFTs, which was discussed in \cite{Shenker:2014cwa,Chen:2016cms,Cotler:2018zff,Haehl:2018izb}, but never in the present language as far as we are aware.

\subsection{Schwarzian theory}
\label{sec:schwarzian}

Consider the Schwarzian theory, described by the action in Lorentzian time $t$, 
\begin{equation}
  iS = - iC \int dt \; \{ f(t) , t \} 
\end{equation}
We now place this action on the contour as in figure \ref{fig:SK}, parameterized by contour time $s$ and corresponding complex time values $t_I(s)$. Our goal is to compute an out-of-time-order four-point function in the configuration \eqref{eq:OTOCconfig}. 
The action has an $SL(2,\mathbb{R})$ symmetry and the $SL(2,\mathbb{R})$ invariant thermal saddle point solution is
\begin{equation}
  f_{I}(s) = \tanh \left( \frac{t_I(s)}{2} \right) \qquad (s\in I)\,.
\end{equation}
where we set $\beta=2\pi$ for simplicity, and $I=1,\ldots,5$ labels the contour segments.
We now consider fluctuations $\epsilon_I(t)$ around this saddle point by sending
\begin{equation}
  f_I(s) \,\longrightarrow\, \tanh \left( \frac{t_I(s)+\de \epsilon_I(t(s))}{2} \right)\,.
\end{equation}
The quadratic action of fluctuations is
\begin{equation}
\label{eq:SchwQuad}
 iS_{quad} = \frac{iC}{2} \int_{\cal C} dt \, \delta\epsilon_I \, (\partial_t^4 - \partial_t^2)\, \delta\epsilon_I 
  = \int_0^{4T+\beta} \!\!ds \, r_s \, \delta \epsilon_I \, (\partial_s^4 + r_s^2 \partial_s^2 ) \, \delta \epsilon_I \,,
\end{equation}
where $r_s$ keeps track of the contour orientation:
\begin{equation}
\label{eq:rsDef}
r_s\equiv i {dt_I (s)\ov ds}=\begin{cases}i \qquad\quad \text{(}I=1,3\text{: `forwards')} \\ -i \qquad\; \text{(}I=2,4\text{: `backwards')}\\1  \qquad\quad \text{(}I=5\text{: `Euclidean')}\end{cases}
\end{equation}
The saddle point reparametrization $f_I(s)$ generates the thermal two-point function for contour-ordered dimension-$\Delta$ operators:
\begin{equation}
\label{eq:GDeltaDef}
\twopt^{\Delta}_{IJ}(s_1,s_2)
 = \left[ \frac{1}{r_{s_1}r_{s_2}} \frac{f_I'(s_1) f_J'(s_2)}{\left( f_I(s_1)-f_J(s_2) \right)^2} \right]^\Delta \qquad (s_1 \in I\,,\; s_2 \in J)\,.
\end{equation}

\begin{figure}
 \begin{center}
       \includegraphics[width=0.65\linewidth]{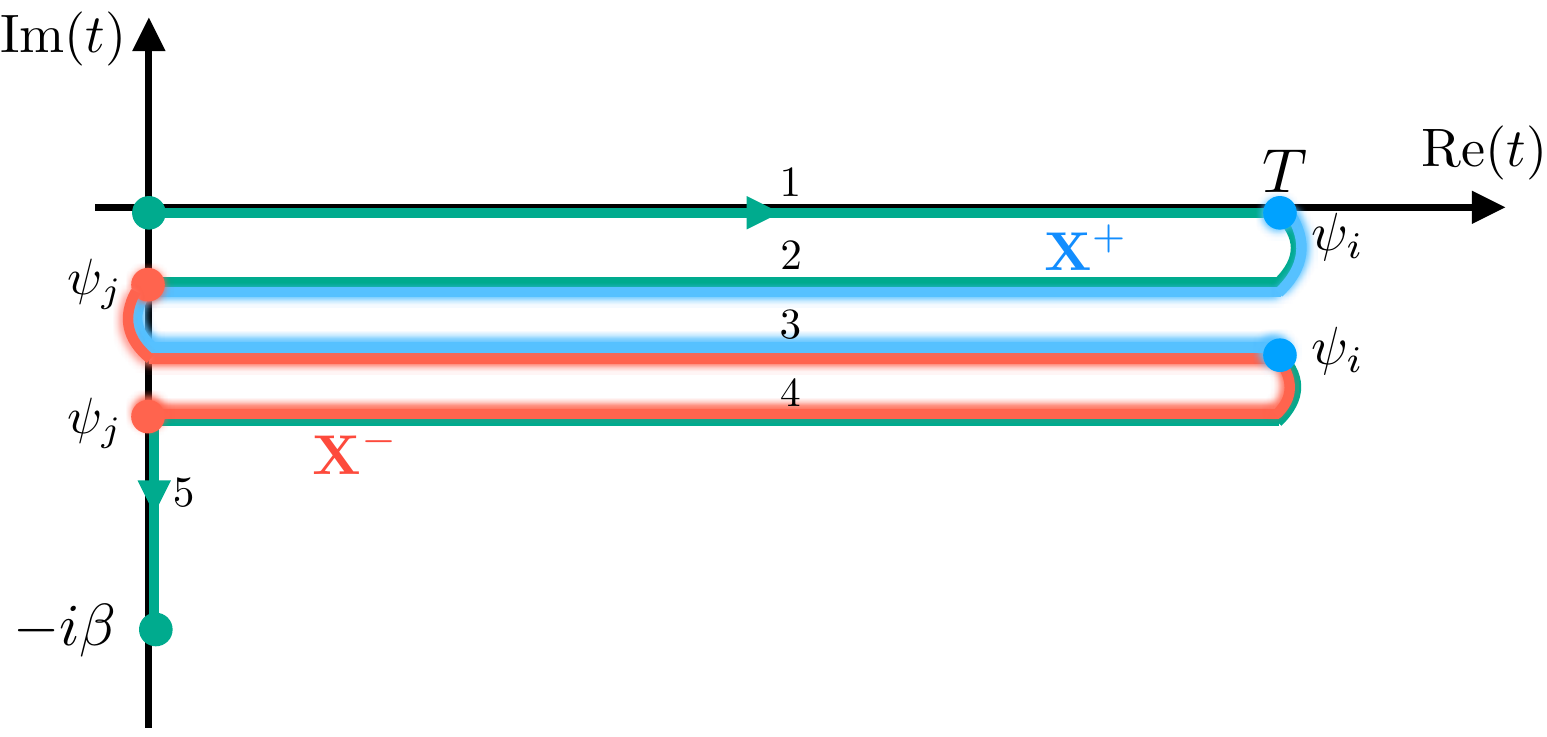}
      \end{center}
     \caption{The support along the SK contour of forward and backward shocks, sourced by pairwise operator insertions at $\hat{t}_1=\hat{t}_2=0$ and $\hat{t}_3 = \hat{t}_4=T$.}
     \label{fig:shocks}
\end{figure}

The action $S_{quad}$ has the usual exponentially growing zero modes:
\begin{equation}
\label{eq:epsOTOC0}
\delta_+ \epsilon_I = X^+ e^{-t(s)}\,,\qquad  \delta_- \epsilon_I = X^- \,e^{t(s)-T}\,.
\end{equation}
 These modes are responsible for the exponential decay of the OTOC (quantum butterfly effect) since they lead to large effects in the future (past) time even if they are excited very weakly in the past (future). We excite them according to
\begin{equation}
\label{eq:epsOTOC}
  \delta \epsilon_I^\text{otoc} = (\chi_{I2} + \chi_{I3}) \, \delta_+ \epsilon_I + (\chi_{I3} + \chi_{I4}) \, \delta_- \epsilon_I \,,
\end{equation}
where $\chi_{IJ}$ is the characteristic function on contour $I=J$, for example: $\chi_{I2} = \Theta(s-T) - \Theta(s-2T)$. We illustrate this pattern in figure \ref{fig:shocks}.
 These modes are not quite continuous across the turning points for finite $T$; they have discontinuities of the size $e^{-T}$, and we would have to find a configuration of the $\epsilon_I$ that is close to the one in \eqref{eq:epsOTOC}, but is continuous. This can be done in an expansion for large $T$. In the end we only need the action of these modes. Similarly to how first order energy shifts in quantum mechanics can be evaluated using the unperturbed eigenfunctions, we can obtain the action using the discontinuous field configuration \eqref{eq:epsOTOC}. 
In Appendix~\ref{app:Brownian} in a related context we provide much more details about this subtlety, but in the main text we will content ourselves with using discontinuous field configurations to obtain their action.

The bulk contribution to the quadratic action \eqref{eq:SchwQuad} is zero when evaluated on this mode. However, the action does receive contributions at the contour turning points $t=0$ and $t=T$ due to the discontinuities of $\delta \epsilon_I^\text{otoc}$ at these locations. Specifically, we find boundary terms when the $s$-derivatives act on the characteristic functions, for example: $\partial_s \chi_{I2}  = \delta(s-2T) - \delta(s-3T)$. Keeping track of such boundary terms we find by evaluating \eqref{eq:SchwQuad} on \eqref{eq:epsOTOC}:\footnote{Note that we do not need to keep higher order terms in $X^+ X^-$ in $S_{eik}$: these would be of the same order in $C$, but involve higher powers of $e^{-T}$. They generate diagrams with higher-point interactions between $X^+$ and $X^-$, which are suppressed compared to diagrams proportional to $(e^T/C)^n \sim {\cal O}(1)$.}
\begin{equation}
\begin{split}
 iS_{eik} \equiv iS_{quad}[\delta \epsilon_I^\text{otoc}] &= \int ds \; r_s\,\partial_s^2 \big(\delta \epsilon_I^\text{otoc} \big) \, (\partial_s^2 + r_s^2 )\delta \epsilon_I^\text{otoc}
 =  2iC \, e^{-T} \, X^+ X^- \,.
 \end{split}
\end{equation}

Natural operators that couple via this action are the bilocal operators obtained by reparameterizing conformal two-point functions \eqref{eq:GDeltaDef}. To this end, consider the finite piecewise $SL(2,\mathbb{R})$ transformations corresponding to \eqref{eq:epsOTOC}:
\begin{equation}
\begin{split}
\label{eq:fSchwFinite}
 \tilde{f}^+_I(s) &\equiv f_{I}(s) + (\chi_{I2} + \chi_{I3}) \, \frac{(1-f_{I}(s))^2 X^+}{2+(1-f_{I}(s))X^+} \,,\\
 \tilde{f}^-_I(s) &\equiv f_I(s) - (\chi_{I3} + \chi_{I4}) \,\frac{(1+f_{I}(s))^2 \,e^{-T}X^-}{2+(1+f_{I}(s))\,e^{-T}X^-}  \,.
\end{split}
\end{equation}
The arrangement of operators as in figure \ref{fig:shocks} corresponds to the following vertex functions:
\begin{equation}
\begin{split}
\twopt^\Delta_{42+}(\hat{s}_1,\hat{s}_2) 
&= \left[ -\frac{\partial f_4(s_1) \,\partial \tilde{f}^+_2(s_2)}{\left( f_4(s_1)-\tilde{f}^+_2(s_2) \right)^2} \right]^\Delta
 = \left[ \frac{1}{1 - \frac{ie^{2i\delta}}{2\sin \delta} \, X^+ } \right]^{2\Delta} \twopt^\Delta_{42}(\hat{s}_1,\hat{s}_2)\,,\\
\twopt^\Delta_{31_-}(\hat{s}_3,\hat{s}_4) &= \left[ - \frac{\partial \tilde{f}^-_3(s_1) \,\partial f_1(s_2)}{\left( \tilde{f}^-_3(s_1)-f_1(s_2) \right)^2} \right]^\Delta= \left[ \frac{1}{1 - \frac{ie^{-i\delta}}{2\sin \delta} \, X^- } \right]^{2\Delta} \twopt^\Delta_{31}(\hat{s}_3,\hat{s}_4)\,,
\end{split}
\end{equation}
where $\delta$ is the contour separation and serves as a regulator (see figure \ref{fig:SK}).

The dominant late-time contribution to the four-point function is now simply given by performing the path integral over the `scramblon' modes $X^\pm$, while ignoring any other (not exponentially growing) modes: 
\begin{equation}
\text{OTOC}_{eik}  = \int {dX^+ dX^-\ov {\cal Z}} \;   \frac{\twopt^\Delta_{42+}(\hat{s}_1,\hat{s}_2)}{\twopt^\Delta_{42}(\hat{s}_1,\hat{s}_2)}   \,\frac{\twopt^\Delta_{31-}(\hat{s}_3,\hat{s}_4)}{\twopt^\Delta_{31}(\hat{s}_3,\hat{s}_4)}  \, e^{i S_{eik}[X^+,X^-]}
 =  z^{-2\Delta} \, U(2\Delta,1,z^{-1})
\end{equation}
where the `cross ratio'  $z=\frac{i\, e^{T}}{8 C \,\sin^2\delta}$ and  ${\cal Z}$ is a measure factor chosen such that a canonical normalization is obtained (i.e., the integral is 1 to leading order in the large $N$ expansion). This result was obtained using very similar considerations in \cite{Maldacena:2016upp}.

\subsection[Two-dimensional CFTs]{Two-dimensional CFTs\protect\footnote{We thank Ying Zhao for discussions on this section.}}

To discuss a second maximally chaotic example, consider the vacuum block contribution to the four-point function in a two-dimensional CFT. Again, one can formulate a theory of reparametrizations, which naturally couple to stress tensors and encode the identity conformal block. The reparametrization mode action arises due to the spontaneous breaking of Virasoro symmetry by the vacuum (or the thermal state) and the conformal anomaly \cite{Haehl:2018izb,Turiaci:2016cvo,Haehl:2019eae,Nguyen:2022xsw}. It also arises as the action for `boundary gravitons' in the path integral quantization of AdS$_3$ gravity, as well as the quantization of the Virasoro coadjoint orbit $\text{diff}(S^1)/\text{PSL}(2,\mathbb{R})$ 
\cite{Polyakov:1987zb,Alekseev:1988ce,Witten:1988,Cotler:2018zff}. The non-linear action can be found in these references. We only require the quadratic action for small reparametrizations $\delta \epsilon_I$, which reads as follows when written in terms of contour time:
\begin{equation}
\label{eq:SCFT}
 iS_{quad} = \frac{c}{24\pi} \int ds d\sigma\!  \left[ \frac{1}{2} (\partial_s + ir_s \partial_\sigma) \delta \epsilon_I \;(r_s \partial_s^3 + r_s^3 \partial_s) \delta \epsilon_I 
 +
  \frac{1}{2} (\partial_s -ir_s \partial_\sigma)  \delta \bar{\epsilon}_I \; (r_s\partial_s^3 + r_s^3 \partial_s) \delta \bar{\epsilon}_I\right] 
\end{equation}
where the complex time coordinate $t\equiv t(s)$ and the spatial coordinate $\sigma$ are related to the holomorphic coordinates of the plane via $(z,\bar{z}) = (e^{t+\sigma} , e^{-t+\sigma})$.\footnote{We are studying the theory on the cylinder: on the infinite spatial line at finite temperature with the choice $\beta=2\pi$. This explains the conformal mapping we employ.} The derivatives $\frac{1}{2}(\partial_s \pm i r_s \partial_\sigma)$ are contour-time versions of (anti-)holomorphic derivatives.

The action $S_{quad}$ has two copies of space-dependent $SL(2,\mathbb{R})$ symmetries, one for each of the terms in \eqref{eq:SCFT}. We will focus on the `holomorphic' first term, as it can be treated as independent of the second copy. The zero modes are analogous to the Schwarzian theory, but importantly they have arbitrary space-dependence: 
\begin{equation}
\label{eq:epsOTOCcft}
\begin{split}
\delta_+ \epsilon_I &= X^+(\sigma') \,e^{-t(s)} \,,\qquad  \delta_- \epsilon_I = X^-(\sigma') \,e^{t(s)-T} \,,\\
\delta_+\bar{\epsilon}_I &= \bar{X}^+(\sigma') \,e^{-t(s)} \,,\qquad  \delta_- \bar{\epsilon}_I = \bar{X}^-(\sigma') \,e^{t(s)-T} \,.
\end{split}
\end{equation}
Considering again the combination of $\delta_\pm \epsilon_I$ appropriate for the OTOC calculation (and similarly for $\delta_\pm \bar{\epsilon}_I$), as in \eqref{eq:epsOTOC}, we find by explicit evaluation of the quadratic action:
\begin{equation}
 iS_{eik} 
= -\frac{ic}{6}\, e^{-T} \int dp\, \left[ (1 + ip) X^+(p) X^-(p) + (1 - ip) \bar{X}^+(p) \bar{X}^-(p) \right]
\end{equation}
where we decomposed into momentum modes, $X^\pm(\sigma') = \int dp \, e^{ip\sigma'} X^\pm(p)$. Note the zeros at $p= \pm i$.

Bilocal operators are again obtained by reparameterizing conformal two-point functions. The relevant reparametrization consists of (two copies of) finite $SL(2,\mathbb{R})$ transformation whose infinitesimal form is \eqref{eq:epsOTOCcft}. This is analogous to (two copies of) the expression \eqref{eq:fSchwFinite}.
The bilocal couplings to the scramblon modes for operators with conformal weights $(h,\bar{h})$ and $(h',\bar{h}')$ are:
\begin{equation}
\begin{split}
\frac{\twopt^{h,\bar{h}}_{42+}(\hat{s}_1,\hat{\sigma}_1;\hat{s}_2,\hat{\sigma}_2)}{\twopt^{h,\bar{h}}_{42}(\hat{s}_1,\hat{\sigma}_1;\hat{s}_2,\hat{\sigma}_2)}&= \left( \frac{1}{1 - \frac{ie^{2i\delta}}{2\sin \delta} \, X^+(\sigma) } \right)^{2h} \left( \frac{1}{1 - \frac{ie^{2i\delta}}{2\sin \delta} \, \bar{X}^+(\sigma) } \right)^{2\bar h}  \,,\\
\frac{\twopt^{h',\bar{h}'}_{31-}(\hat{s}_3,\hat{\sigma}_3;\hat{s}_4,\hat{\sigma}_4)}{\twopt^{h',\bar{h}'}_{31}(\hat{s}_3,\hat{\sigma}_3;\hat{s}_4,\hat{\sigma}_4)}& = \left( \frac{1}{1 - \frac{ie^{-i\delta}}{2\sin \delta} \, X^-(\sigma') } \right)^{2h'}  \left( \frac{1}{1 - \frac{ie^{-i\delta}}{2\sin \delta} \, \bar{X}^-(\sigma') } \right)^{2\bar h'} \,,
\end{split}
\end{equation}
where $\sigma$ and $\sigma'$ are the spatial insertions of the operators in the future (at $t=T$) and in the past (at $t=0$), respectively: $\hat{\sigma}_1 = \hat{\sigma}_2 = \sigma$, $\hat{\sigma}_3 = \hat{\sigma}_4 = \sigma'$. It is useful to write a `bulk' representation of these vertex functions in terms of lightcone momentum wave functions \cite{Maldacena:2016upp}. For example, the holomorphic parts of the bilocal operators can be written as:
\begin{equation}
\label{eq:holoIntegral}
\begin{split}
\left( \frac{1}{1 - \frac{ie^{2i\delta}}{2\sin \delta} \, X^+(\sigma) } \right)^{2h}
&= \frac{1}{\Gamma(2h) (2\sin\delta)^{2h}}\int_{-\infty}^0 dq_+ \, \frac{1}{-q_+} \Psi_1^h(q_+) \Psi_2^h(q_+) \,e^{ - iq_+ X^+(\sigma) } \\
\left( \frac{1}{1 - \frac{ie^{-i\delta}}{2\sin \delta} \, X^-(\sigma') } \right)^{2h'}
&= \frac{1}{\Gamma(2h')(2\sin\delta)^{2h'}}\int_{-\infty}^0 dp_- \, \frac{1}{-p_-} \Psi_3^{h'}(p_-) \Psi_4^{h'}(p_-) \, e^{-i p_-  X^-(\sigma')} 
\end{split}
\end{equation} 
where $\Psi_j^h \equiv (x_j)^{h} \, e^{-x_j}$ with
\begin{equation}
 x_1 = -iq_+ e^{-3i\delta}  \,,\quad x_2 = i q_+ e^{-i\delta}  \,,\quad x_3 = i p_- e^{2i\delta}\,,\quad x_4 = -i p_-  \,.
\end{equation}
We can then write the holomorphic part of the eikonal four-point function as follows:
{\small
\begin{equation}
\begin{split}
& \text{OTOC}^{(holo)}_{eik}\big(\hat{t}_1+\hat{\sigma}_1,\hat{t}_2+\hat{\sigma}_2,\hat{t}_3+\hat{\sigma}_3,\hat{t}_4+\hat{\sigma}_4\big)  = \\
&\;\; =  \frac{(2\sin\delta)^{-2(h+h')}}{\Gamma(2h)\Gamma(2h')}\int {dX^+ dX^-\ov {\cal Z}} 
\int_{-\infty}^0 dq_+ dp_-  \le[ { \Psi_1^h(q_+) \Psi_2^h(q_+) \ov -q_+} \ri] \le[ { \Psi_3^{h'}(p_-) \Psi_4^{h'}(p_-) \ov -p_-} \ri] \\
 &\qquad\qquad\quad\times \exp \left\{\int dp \left[ \left(-\frac{ic}{6} \, e^{-T} (1+ip) \right) X^+(p) X^-(p) -iq_+  e^{ip\sigma} X^+(p)- i p_-  e^{ip\sigma'}X^-(p)\right]\right\} \\
 &\;\;= \frac{(2\sin\delta)^{-2(h+h')}}{\Gamma(2h)\Gamma(2h')}\int_{-\infty}^0 dq_+ dp_-\!  \le[ { \Psi_1^h(q_+) \Psi_2^h(q_+) \ov -q_+} \ri] \le[ { \Psi_3^{h'}(p_-) \Psi_4^{h'}(p_-) \ov -p_-} \ri] \exp \left\{ \frac{6i}{c} \int dp \, \frac{q_+ p_-}{1+ip} \, e^{T+ip(\sigma'-\sigma)} \right\}
\end{split}
\end{equation}
}\normalsize
The integral in the exponential is determined by a `graviton' pole at $p=i$, and evaluation per residue gives:
{\small
\begin{equation}
\begin{split}
 &\text{OTOC}^{(holo)}_{eik}  =\\
  &\;\; = \frac{(2\sin\delta)^{-2(h+h')}}{\Gamma(2h)\Gamma(2h')}\int_{-\infty}^0 dq_+ dp_-\!  \le[ { \Psi_1^h(q_+) \Psi_2^h(q_+) \ov -q_+} \ri] \le[ { \Psi_3^{h'}(p_-) \Psi_4^{h'}(p_-) \ov -p_-} \ri] \exp \left\{ \frac{12\pi i}{c} \, q_+ p_- \,\Theta(\sigma'-\sigma) \, e^{T-(\sigma'-\sigma)}  \right\}
\end{split}
\end{equation}
}\normalsize
A similar expression applies for the anti-holomorphic part of the four-point function, which only depends on the combinations $\hat{t}_i - \hat{\sigma}_i$. It will involve an exponent proportional to $\Theta(\sigma-\sigma') \,e^{T-(\sigma-\sigma')}$. The full four-point function consists of the product of these two expressions and evaluates to:
\begin{equation}
\label{eq:CFTresult}
 \text{OTOC}^{(holo)}_{eik} \, \overline{\text{OTOC}^{(holo)}_{eik}}
 =  z^{-2h'} U\left(2h',1+2h'-2h,\frac{1}{z}\right) \, \times \, \bar{z}^{-2\bar{h}'} U\left(2\bar{h}',1+2\bar{h}'-2\bar{h},\frac{1}{\bar{z}}\right)
 \end{equation}
 with 
 \begin{equation}
 z \equiv -\frac{12\pi i}{c} \, (2\sin\delta)^{-2}\, \Theta(\sigma'-\sigma) \, e^{T+(\sigma-\sigma')} \,,\quad\;\;\;
 \bar z \equiv -\frac{12\pi i}{c} \, (2\sin\delta)^{-2}\, \Theta(\sigma-\sigma') \, e^{T-(\sigma-\sigma')}\,.
\end{equation}
The holomorphic and anti-holomorphic factors in \eqref{eq:CFTresult} become 1 whenever the spatial step functions vanish. The presence of the step functions can be understood by keeping in mind that we only keep track of terms that grow exponentially as $T \rightarrow \infty$. We would obtain terms with step functions of opposite support if we had considered $T \rightarrow -\infty$. In other words: the only exponentially growing term that is {\it holomorphic} has to depend on $T+(\sigma-\sigma')$, which only exhibits exponential decay in space as long as $\sigma-\sigma' < 0$. In \cite{Chen:2016cms}, equivalent expressions for analytic continuations of identity conformal blocks were obtained using Lorentzian resummation methods (see also \cite{Lam:2018pvp}).

\section{The large $q$ SYK model}

A key question in the study of scrambling in large $N$ systems is how to go away from the best understood maximally chaotic case. While we currently lack the generalization of the Schwarzian action to the case of sub-maximal chaos, we find that the $X^\pm$ scramblon modes can be adapted to this case as well.
Having understood the basic strategy in maximally chaotic systems, we take inspiration from our previous calculations and we shall now adapt this strategy in the large $q$ SYK model, where the Lyapunov exponent is less than maximal.

\subsection{Overview}

The large $q$ limit of SYK was introduced in \cite{Maldacena:2016upp}: starting from the SYK Hamiltonian, 
\begin{equation} 
 H = (i)^{\frac{q}{2}} \sum_{1 \leq i_1 < \cdots < i_q \leq N} \, J_{i_1 \cdots i_q} \, \chi_{i_1} \cdots  \chi_{i_q} \,,\qquad\;\;\;  \left\langle J_{i_1 \cdots i_q}^2 \right\rangle = \frac{(q-1)!}{N^{q-1}} \, J^2 \equiv \frac{2^{q-1}(q-1)!}{q\, N^{q-1}} \, {\cal J}^2 \,,
\end{equation}
one takes the limit $N \gg q^2 \gg 1$, while holding ${\cal J}^2$ fixed.\footnote{ A different large $q$ limit was explored in \cite{Berkooz:2018qkz,Berkooz:2018jqr}. It would be interesting to study it using the methods we develop here.} One  expands the usual collective $(\twopt,\Sigma)$ variables as
\begin{equation}
 \twopt(\tau,\tau') = \frac{\text{sgn}(\tau-\tau')}{2} \left[ 1 + \frac{g(\tau,\tau')}{q} + \ldots \right] \,,\qquad \Sigma(\tau,\tau') = \frac{{\cal J}^2}{q}\, \text{sgn}(\tau-\tau') \, e^{g(\tau,\tau')}\left[ 1 + \ldots \right] \,,
\end{equation}
and discovers a non-trivial interacting theory at ${\cal O}(1/q^2)$. Indeed, the large $N$ effective action to leading non-trivial order in the $1/q$ expansion takes the form of a Liouville-type theory; placed on the generalized Schwinger-Keldysh contour ${\cal C}$, it takes the following form:
\beq
\label{eq:SLiouville}
iS[g]={N\ov 4q^2}\int  ds_1 ds_2 \left[ {1\ov 4}\pa_1 g\pa_2 g-\mathcal J^2  r_{s_1} r_{s_2}\,e^g\right] \,,
\eeq
with $r_s \equiv i \frac{dt(s)}{ds}$ as in \eqref{eq:rsDef}.
The associated equation of motion is a Liouville equation, whose solutions we will describe in more detail in the next subsection.
Importantly, the field $g(s_1,s_2)$ obeys a symmetry condition $g(s_1,s_2)=g(s_2,s_1)$ and the KMS conditions $g(s_1,4T+2\pi)=g(s_1,0),~g(4T+2\pi,s_2)=g(0,s_2)$.

In the large $q$ SYK model, the result for the four-point function \eqref{eq:OTOCconfigDef} at first non-trivial order in $1/N$ for the arrangement \eqref{eq:OTOCconfig} is known exactly \cite{Streicher:2019wek,Choi:2019bmd}:\footnote{ The OTOC in refs.\ \cite{Streicher:2019wek} and \cite{Choi:2019bmd} differs by a factor of $\frac{1}{4}$ due to the use of different normalizations for the SYK fermion operators. We use the conventions of the latter, i.e., $\chi_i^2 = \frac{1}{2}$. Note, however, that in \eqref{eq:FOTOClargeq} we removed the normalization factor $\frac{1}{4}$ thus obtained, as we defined ${\cal F}$ in \eqref{eq:OTOCconfigDef} to be normalized by two-point functions, which differs from the definition of ${\cal F}$ in \cite{Choi:2019bmd}.\label{foot:normalization}} 
\begin{equation}
\label{eq:FOTOClargeq}
\begin{split}
   {\cal F}^\text{otoc}(t_1,t_2,t_3,t_4) &= {\cal F}_{exp}(t_1,t_2,t_3,t_4) + {\cal F}_{poly}(t_1,t_2,t_3,t_4) \,,\\
   {\cal F}_{exp} &= - \frac{2\cosh \left( \frac{v}{2}(i\pi + t_1+t_2 -t_3-t_4) \right)}{\cos \left( \frac{\pi v}{2} \right)\cosh \left( \frac{v}{2} \,(t_{12}+i\pi) \right) \cosh \left( \frac{v}{2} \, (t_{34} +i \pi) \right) } \,,\\
   {\cal F}_{poly} &= \frac{2\tan \left( \frac{\pi v}{2} \right)}{\frac{\pi v}{2} +\cot \left( \frac{\pi v}{2} \right)} \left(1- \frac{v(i\pi+t_{12})}{2\coth \left( \frac{v}{2} (i\pi +t_{12}) \right)  } \right) \left(1- \frac{v(i\pi+t_{34})}{2 \coth\left( \frac{v}{2} (i\pi +t_{34}) \right)  } \right) \\
     &\qquad - \tan \left( \frac{\pi v}{2} \right) \, \frac{ iv(i\pi +t_1+t_2-t_3-t_4) }{\coth \left( \frac{v}{2} (i\pi +t_{12}) \right) \coth \left( \frac{v}{2} (i\pi +t_{34}) \right)} \,,
 \end{split}
\end{equation}
where we introduced $v$ as a proxy for ${\cal J}$ through the equation ${\cal J}={v/(2\cos(\pi v/2))}$.
Specifically, for our choice of configuration \eqref{eq:OTOCconfig}, it exhibits the following large $T$ behavior:
\beq
\begin{split}
 {\cal F}^\text{otoc}(\hat{t}_1,\hat{t}_2,\hat{t}_3,\hat{t}_4) 
 & = -  e^{-\frac{i\pi v}{2}} \, \sec \left( \frac{\pi v}{2} \right)^{3} \, e^{vT} - \frac{iv}{2} \tan \left( \frac{\pi v}{2} \right)^3 \, T + {\cal O}\left( (e^{vT})^0\right)
 \end{split}
 \label{eq:OTOCres}
\eeq
where the last line gives the asymptotics in a large $T$ expansion with all higher order corrections being of the form $e^{-nvT}$ with $n=0,1,2,\ldots$. We emphasize that our approach straightforwardly reproduces the dependence of the OTOC on $t_{12}$ and $t_{34}$, and we have demonstrated this explicitly. Here we are studying the simplest configuration of times for presentation purposes. 

We will develop an effective description of the large $T$ behavior of the OTOC, using similar methods as in the case of maximal chaos. The theory is based on the large $q$ effective action on the Schwinger-Keldysh contour, \eqref{eq:SLiouville}.

\subsection{Thermal solution}

Our starting point is the large $q$ effective action \eqref{eq:SLiouville} on the Schwinger-Keldysh contour. The equation of motion is
\beq
 \frac{1}{2} \partial_1 \partial_2 g(s_1,s_2) + {\cal J}^2 r_{s_1} r_{s_2} \, e^{g(s_1,s_2)} = 0 \,.
\eeq
While $g(s_1,s_2)$ is a single-valued, continuous field on the torus $[0,4T+2\pi]^2$, it is convenient to introduce the notation $g_{IJ}$ as a restriction of $g$ defined on the region $s\in I, s'\in J$ where $I,J=1,\dots, 5$ are the labels of each segment on the SK contour shown in figure \ref{fig:SK}.

The thermal saddle point solution can be constructed as follows. First, since $r_s$ is constant along the each contour segment, we can use the most general vacuum solution of the Liouville equation to parametrize each $g_{IJ}$ separately as follows:
\beq
e^{g_{IJ}(s_1,s_2)}={1\ov \mathcal J^2 r_{s_1} r_{s_2}}{F_{IJ}'(s_1) G_{IJ}'(s_2)\ov (F_{IJ}(s_1)-G_{IJ}(s_2))^2} \qquad (s_1\in I, s_2\in J)
\eeq
These solutions are invariant under an $SL(2,\mathbb{R})_{diag}$ symmetry
\beq
  F_{IJ}(s) \; \longrightarrow \; \frac{a\, F_{IJ}(s) +b}{c\, F_{IJ}(s) + d}\,,\qquad G_{IJ}(s) \; \longrightarrow \; \frac{a \,G_{IJ}(s) +b}{c \,G_{IJ}(s) + d} \,,\qquad ad-bc=1\,.
\eeq
Note that the Liouville action had a larger $\text{diff}(S^1) \times \text{diff}(S^1)$ symmetry, of which only $SL(2,\mathbb{R})_{diag}$ is preserved by the solution. We give some more details about this symmetry in appendix \ref{sec:SL2} (see also \cite{Lin:2022rbf,StanfordTalk} for a more thorough investigation).

The remaining task is to find $(F_{IJ},G_{IJ})$ such that the solution respects the symmetry and the boundary condition of $g_{IJ}$. First, the symmetry $g(s_1,s_2)=g(s_2,s_1)$ can be achieved straightforwardly if we restrict to $s_1>s_2$ (and hence $I\geq J$) and extend the solution to the other half using the desired symmetry. We find that the general solutions for $F_I,G_I$ in terms of the complex time $t(s)$ are given by (up to $SL(2,\mathbb R)_{diag}$ transformations of $(F_{IJ},G_{IJ})$):
\beq
~&F_{IJ}(s)=\tanh( {v t_I(s)+2\pi v i \ov 2}+c_{IJ}) \,,
\\& G_{IJ}(s)=\tanh( {v t_J(s)+\pi (1+v)i\ov 2}+c_{IJ}) \,,
\eeq
where $c_{IJ}$ is an arbitrary constant, which depends only on the associated region $(I,J)$. Although $g_{IJ}$ is independent of the shifts $c_{IJ}$, the latter play an important role in the following discussion.

\subsection{Quadratic action of fluctuations}
\label{sec:actionCalc}

To analyze the eikonal action in the $T\rightarrow\infty$ limit, we need to find near-zero eigenmodes around the vacuum $g_{IJ,vac}$ constructed from above $F_{IJ},G_{IJ}$. Semiclassically, we have to analyze the following quadratic action
\beq
iS_{quad}~&= {N\ov 4q^2}\iint  ds_1 ds_2  \,\del g \left(  - {1\ov 4}\pa_1 \pa_2-{1\ov 2}\mathcal J^2  r_{s_1} r_{s_2}\,e^g\right)\del g
\\&\equiv {N\ov 4q^2}\ex{\del g, \mathcal L \del g}\,,
\eeq
where in the second line we defined a Liouville operator  $\mathcal L=\left( - {1\ov 4}\pa_1 \pa_2-{1\ov 2}\mathcal J^2  r_{s_1} r_{s_2}\,e^g\right)$ as a self-adjoint operator acting on the space of fluctuations $\del g$ with the real inner product $\ex{\del  g_1,\del g_2}=\iint ds_1ds_2 \, \del g_1(s_1,s_2) \del g_2 (s_1,s_2)$.\footnote{In Euclidean signature $\mathcal L$ is self-adjoint and $\de g$ is real. On the SK contour both of them get analytically continued to complex configurations and ultimately will give us complex results.} 

Analogous to the Schwarzian case discussed in section \ref{sec:schwarzian}, one can guess two near-zero eigenmodes $\del_{\pm}g$ in the eikonal limit, each generated from the $SL(2,\mathbb{R})$ shocks on the segments $I=2,3$ and $I=3,4$:
\begin{equation}
\label{eq:FGnonlinear}
\begin{split}
   F_{IJ} & \;\longrightarrow\; F_{IJ} +  (\chi_{I2}+\chi_{I3})\,\frac{X^+ (1-F_{IJ})^2}{2+X^+(1-F_{IJ})}   - (\chi_{I3}+\chi_{I4}) \, \frac{e^{-vT}X^- (1+F_{IJ})^2}{2+e^{-vT}X^-(1+F_{IJ})} \\
      G_{IJ} & \;\longrightarrow\; G_{IJ} +(\chi_{J2}+\chi_{J3}) \,\frac{X^+ (1-G_{IJ})^2}{2+X^+(1-G_{IJ})} - (\chi_{J3}+\chi_{J4})\, \frac{e^{-vT}X^- (1+G_{IJ})^2}{2+e^{-vT}X^-(1+G_{IJ})}
\end{split}
\end{equation}
We will also use the following notation to single out the two shocks:
{\small
\begin{equation}
\label{eq:FGtildeDef}
\begin{split}
  \tilde{F}^+_{IJ}&= F_{IJ} + (\chi_{I2}+\chi_{I3})\, \frac{X^+ (1-F_{IJ})^2}{2+X^+(1-F_{IJ})} \,,\qquad\;\; \tilde{F}^-_{IJ} = F_{IJ}- (\chi_{I3}+\chi_{I4}) \frac{e^{-vT}X^- (1+F_{IJ})^2}{2+e^{-vT}X^-(1+F_{IJ})}  \,,\\
      \tilde{G}^+_{IJ}&=G_{IJ}+(\chi_{J2}+\chi_{J3})\,\frac{X^+ (1-G_{IJ})^2}{2+X^+(1-G_{IJ})} \,,\qquad \tilde{G}^-_{IJ} =G_{IJ} -(\chi_{J3}+\chi_{J4})\, \frac{e^{-vT}X^- (1+G_{IJ})^2}{2+e^{-vT}X^-(1+G_{IJ})} \,.
\end{split}
\end{equation}
}\normalsize
For the moment, we restrict to the transformations linearized in $X^\pm$: 
\beq
\label{eq:FGlin}
~&\del_+ F_{IJ}(s)={(1-F_{IJ}(s))^2\ov 2}\,(\chi_{I2}+\chi_{I3})
\\&\del_+ G_{IJ}(s)={(1-G_{IJ}(s))^2\ov 2}\,(\chi_{J2}+\chi_{J3})
\\&\del_- F_{IJ}(s)=-{(1+F_{IJ}(s))^2\ov 2}\,e^{-vT}\,(\chi_{I3}+\chi_{I4})
\\&\del_- G_{IJ}(s)=-{(1+G_{IJ}(s))^2\ov 2}\,e^{-vT}\,(\chi_{J3}+\chi_{J4}) 
\eeq
The factors of $e^{-vT}$ in the transformations generating $\delta_- g$ are chosen such that $\delta_- g$ is finite (normalizable) when $t_1\sim t_2\sim T \rightarrow \infty$ (see below, \eqref{eq:norm}).

\begin{figure}
\begin{center}
       \includegraphics[width=0.7\linewidth]{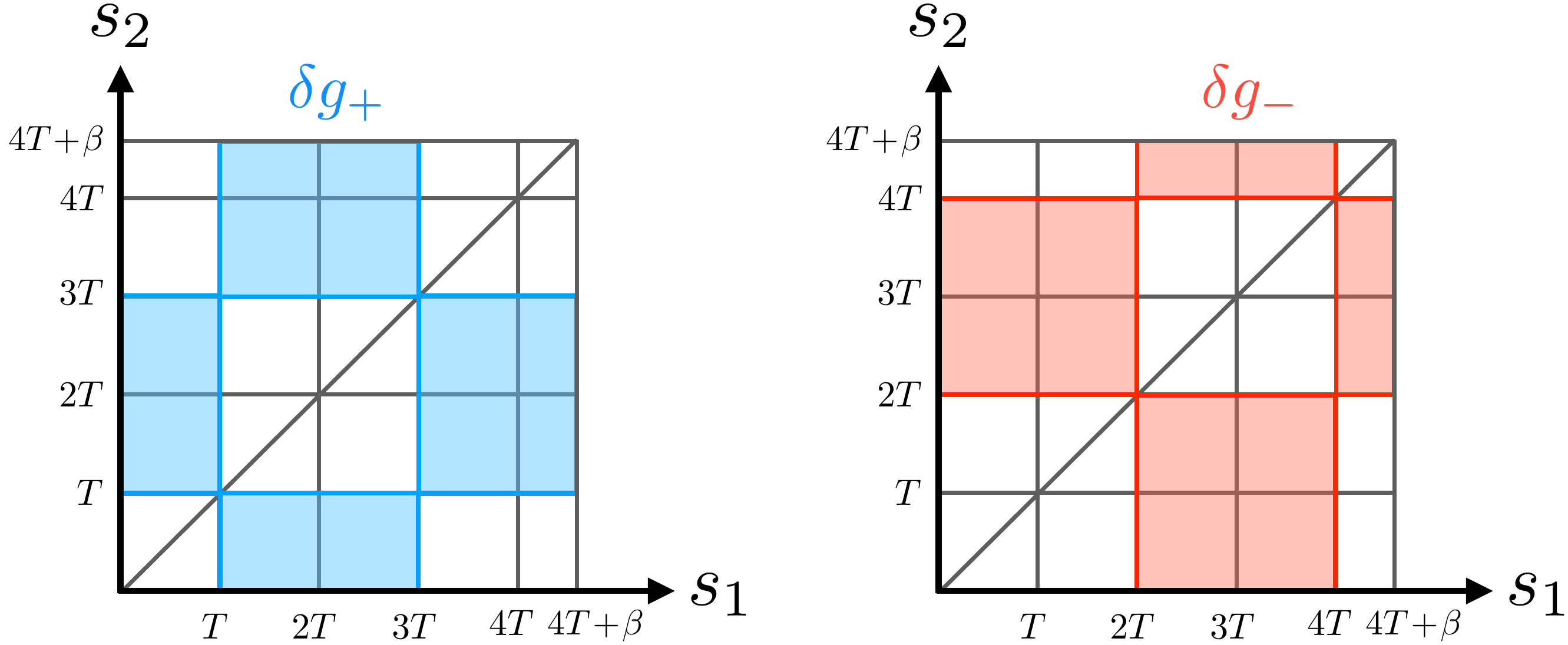}
       \end{center}
     \caption{The bilocal version of figure \ref{fig:shocks}, as required for the large $q$ SYK model. Shaded regions show where the fluctuations $\del_\pm g(s_1,s_2)$ have non-zero support. By symmetry, we can always focus on the triangle $s_1 > s_2$.}
     \label{fig:pmshocks}
\end{figure}

We illustrate the support of the shocks along the contour in figure \ref{fig:shocks}. Note that because of the $SL(2,\mathbb{R})_{diag}$ invariance, there are 12 (out of 25) non-zero regions $(I,J)$ for each $\del_\pm g$ (and hence 6 regions for $s_1>s_2$) as shown in figure \ref{fig:pmshocks}.

 Importantly, $\del_{\pm}g$ derived from the above shock solutions does not obey the KMS condition in general. In order to remedy this, we can tune the $c_{IJ}$ in a specific way for the forward and backward propagating shocks, such that the KMS condition is satisfied. The unique solution up to an overall constant shift in the non-zero regions is given by
\beq
~&c_{+,21}=c_{+,31}={i\pi (1-v)\ov 2}\,,\;~ c_{+,42}=c_{+,52}=c_{+,43}=c_{+,53}=0 \,,
\\&c_{-,53}=c_{-,54}=-{i\pi (1-v)\ov 2}\,,\;~ c_{-,31}=c_{-,41}=c_{-,32}=c_{-,42}=0\,.
\eeq
Physically, the freedom to choose $c_{\pm, IJ}$ is equivalent to the freedom of rescaling $\del_\pm g_{IJ}\rightarrow A_{\pm,IJ} \del_\pm g_{IJ}$ in each region. This understanding leads to the following simple expression for $\del_{\pm}g$ (again for $I\geq J$):
\beq \label{eq:shocksimple}
~&\del_{\pm }g_{IJ}=A_{\pm , IJ}\, \psi^e_v( i(t_1-t_2)) \, e^{\pm v( (1\mp 1)T-t_1-t_2)/2}
\\&A_{+,IJ} =\begin{cases}e^{-{i\pi v\ov 2}} &\quad(I,J)=(2,1),(3,1)
\\
e^{-{3i\pi v\ov 2}}&\quad(I,J)=(4,2),(5,2),(4,3),(5,3)
\\0 &\quad 	\text{otherwise}
\end{cases}, 
\\& A_{-,IJ}=\begin{cases}-e^{{3i\pi v\ov 2}} &\quad(I,J)=(3,1),(4,1),(3,2),(4,2)
\\
-e^{{5i\pi v\ov 2}}&\quad(I,J)=(5,3),(5,4)
\\0 &\quad 	\text{otherwise}
\end{cases}
\eeq
where we are using the notation of \cite{Choi:2019bmd}, i.e., $\psi_v^{e}(x)$ is an even eigenfunction of the Liouville operator ${\cal L}$ of the form 
\beq
\begin{split}
 \psi_v^e(x) &= \frac{1}{\cos \left( \frac{v}{2}(\pi-x) \right)} \,.
\end{split}
\eeq
An important point about \eqref{eq:shocksimple} is that we could have just written it down as an Ansatz from the requirement t near-zero modes and then determined the $A_{\pm,IJ}$ from consistency. Indeed this is the method we will follow in the SYK chain, where there is no apparent symmetry controlling scrambling.

It is straightforward to confirm that $\mathcal L \del_\pm g_{IJ}=0$ away from the boundaries between different regions. Note that in this way of writing the eigenmodes, the factors $e^{-vT}$ in the normalization of $\delta_- g$ follow trivially by demanding that they be finite as $t_1 \sim t_2 \sim T \rightarrow \infty$. Indeed, we find the following expressions for the `norms' of the fluctuations:
\begin{equation}
\label{eq:norm}
\begin{split}
   \langle \delta_+ g, \delta_+ g \rangle &= \frac{16}{v^2} \, e^{-2\pi i v} \left( \cos (\pi v) + \sin (\pi v) \right) \left[  \log \left( 2 \cos \left( \frac{\pi v}{2} \right) \right) \cos (\pi v)- \frac{1-\pi v \sin (\pi v)}{2} \right]+ {\cal O}(e^{-vT})\\
   \langle \delta_- g, \delta_- g \rangle &= \frac{16}{v^2} \, e^{3\pi i v} \left[  \log \left( 2 \cos \left( \frac{\pi v}{2} \right) \right) \cos (\pi v)- \frac{1-\pi v \sin (\pi v)}{2} \right] + {\cal O}(e^{-vT})\,.
\end{split}
\end{equation}
This confirms that the prefactors are chosen such as to ensure normalizability at large $T$. For our later analysis, it will also be important to note that $\text{arg}(\langle \delta_+ g,\delta_+g\rangle\langle \delta_-g,\delta_-g\rangle) = i\pi v$ at large $T$. Relatedly, note that $\delta g_\pm(t_1,t_2)$ is normalized such that $\delta g_+(0,0) = {\cal O}(1)$, $\delta g_+(T,T) = {\cal O}(e^{-vT})$, and vice versa for $\delta g_-$.

\subsection{Derivation of the eikonal action}

The remaining task is to perform a degenerate perturbation theory using $\del _{\pm }g$ and obtain the nearly-zero eigenmodes of $\mathcal L$ and hence the eikonal action. We first emphasize that $\del_{\pm}g$ would have been `exact' zero eigenmodes if they were continuous function on the whole range of $s_1,s_2$. However, the fields $\del_{\pm} g$ are discontinuous along the boundaries between the vanishing and non-vanishing regions (illustrated as colored edges in figure \ref{fig:pmshocks}). Therefore, non-zero contributions to the action are generated when the derivative operator $-\frac{1}{4}\pa_1\pa_2\in \mathcal L$ acts on such configurations and generates boundary terms.\footnote{To handle this technicality, it is convenient to define a characteristic function $\chi_{IJ}(s_1,s_2)$ which is $1$ in the $(I,J)$ region and $0$ outside of the region, and use the relation $\del _{\pm}g=\sum_{I,J}\chi_{IJ}\del_\pm g _{IJ}$.}

We first argue that the `diagonal' contributions vanish, i.e., $\ex{\del_+ g, \mathcal L \del_+ g}=\ex{\del_- g, \mathcal L \del_- g}=0$. This can be understood as follows. There are two classes of non-zero contributions to $\mathcal L \del_\pm g $ coming either from the edges or from the vertices (which are the intersection points of edges). The former (edge) contribution is given by
\beq
\ex{\del g, \pa_1\pa_2 \del g}|_\text{edges}&=\sum_{E\in \,\text{edges}} \int_E ds  \;{1\ov 2}(\del g_{avg})  \pa_s (\del g_{dif\! f})
\\&=\sum_{V\in\, \text{vertices}}{1\ov 2} \left[-(\del g)^2|_{(+,+)}- (\del g)^2|_{(-,-)}+ (\del g)^2|_{(+,-)}+ ( \del g)^2|_{(-,+)}\right]
\eeq
where $\del g_{avg}$ and $ \del g_{dif\! f}$ are averages and differences along the discontinuity of $\del g$ at the edge and $s$ is a canonical integration direction along the edge. Since one side of the discontinuity is always 0, each edge integral reduces to boundary contributions. The $(\pm,\pm)$ notation in the second line denotes the limiting value of $(\del g)^2$ on the vertices approaching from the direction $(\pm,\pm)$.

The vertex contribution comes from the delta-function generated by the action of $\pa_1\pa_2$ on the characteristic function. This directly gives
\beq
\ex{\del g, \pa_1\pa_2 \del g}|_\text{vertices}&=\sum_{V\in \,\text{vertices}} {1\ov 2}\left[ ( \del g)^2|_{(+,+)}+ (\del g)^2|_{(-,-)}- (\del g)^2|_{(+,-)}- ( \del g)^2|_{(-,+)}\right]\,.
\eeq
 Therefore, we see that the contributions from edges and vertices cancel each other for the `diagonal' fluctuations. 
 
\begin{figure}
\begin{center}
       \includegraphics[width=0.38\linewidth]{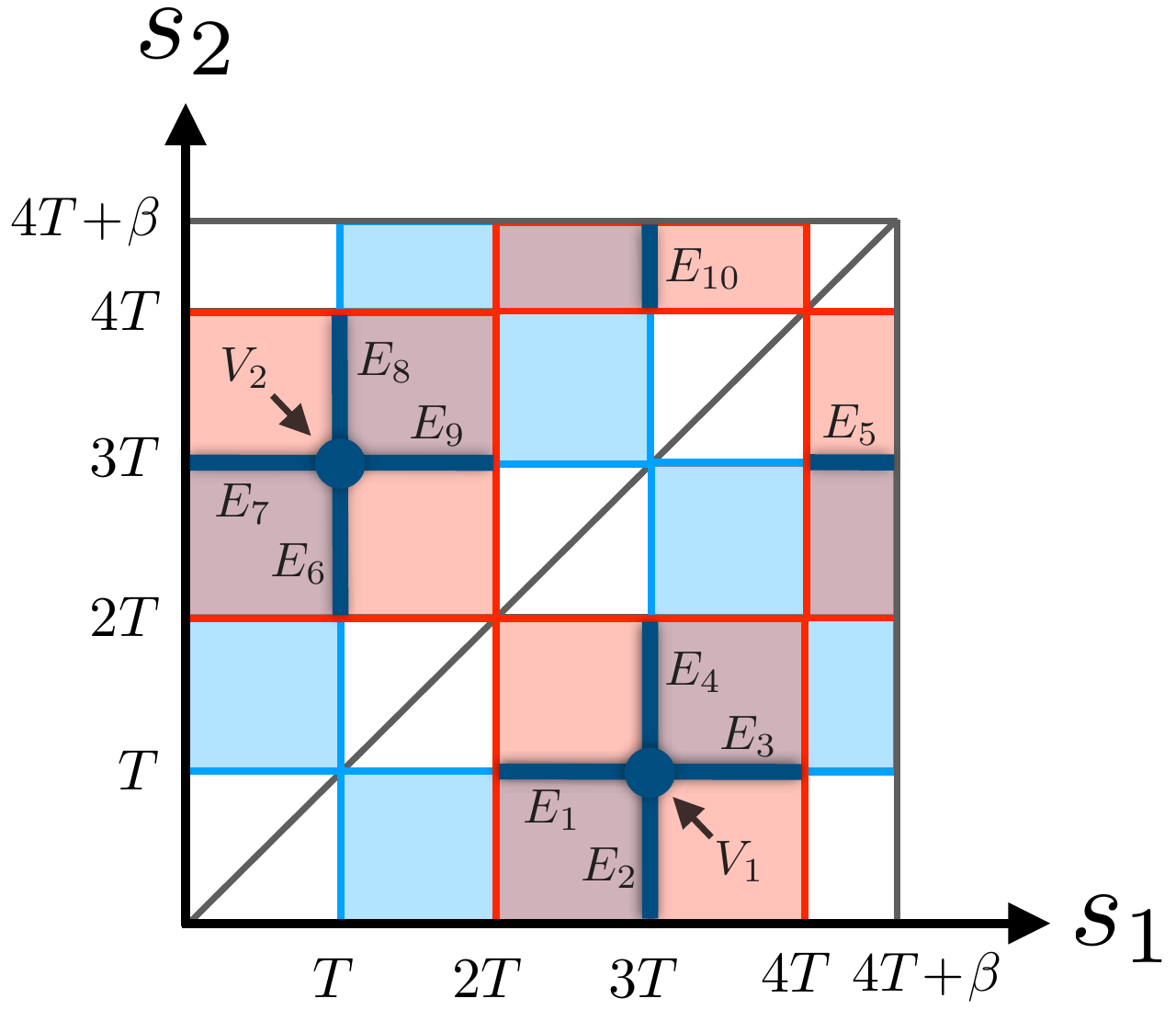}
       \end{center}
     \caption{Edges $E$ and vertices $V$ contributing non-trivial boundary terms to the off-diagonal quadratic action $\ex{\del _-g, \mathcal L \del_+ g}$. A similar picture applies to $\ex{\del _+g, \mathcal L \del_- g}$.}
     \label{fig:off}
\end{figure}

 Next we compute the off-diagonal elements $\ex{\del _-g, \mathcal L \del_+ g}=\ex{\del _+g, \mathcal L \del_- g}$, which will give non-vanishing contributions. As before, the action of $\mathcal L$ on $\del _{\pm} g$ gives edge and vertex contributions, but now they no longer cancel each other. As shown in figure \ref{fig:off}, illustrating the case of $\ex{\del _-g ,\mathcal L \del_+ g}$, there are 10 edges and 2 vertices. Summing up their contributions, we find:
\beq
\ex{\del_- g ,\mathcal L \del_+ g}&=-{1\ov 4} \left(\sum_{E\in \,\text{edges}} \int_E ds \, \del_- g \,  \pa_s \del_+ g_{dif\! f} 
+\sum_{V\in \,\text{vertices}}(\del_- g \, \del _+g|_{(+,+)}+\dots) \right )
\\&=2\cos \left({\pi v \ov 2}\right) e^{{i\pi v\ov 2}-vT}+ O(e^{-2 vT})
\eeq
The contribution from $\ex{\del _+g, \mathcal L \del_- g}$ is identical. This implies that the eikonal action for the overlapping shocks $\del g=X^+\del_+g+X^- \del_- g$ is given by
\beq
\label{eq:Sdot}
\boxed{
iS_{eik}={N\ov q^2}\cos \left({\pi v \ov 2}\right) e^{{i\pi v\ov 2}-vT}\, X^+ X^- 
}
\eeq
This translates into a propagator\footnote{ As in the Schwarzian theory, the primed integral means we choose a measure that achieves a unit normalization of the OTOC at order ${\cal O}(N^0)$. In the present case this means we have a measure factor $\frac{N}{2\pi q^2} \cos\big(\frac{\pi v}{2} \big) e^{\frac{i\pi (v-1)}{2} -vT}$.}
\begin{equation}
\label{eq:Xprop}
 \langle X^+ X^- \rangle \equiv \int {dX^+ dX^-\ov {\cal Z}}  \, (X^+ X^-)\, e^{iS_{eik}} =\frac{q^2}{N} \, \sec \left( \frac{\pi v}{2} \right) e^{-\frac{i\pi v}{2}+vT}   \,.
\end{equation}

\subsection{Bilocal operators and the OTOC}

Having found the eikonal action, we can now confirm that we reproduce the OTOC \eqref{eq:OTOCres} in the long time limit. The missing ingredient for the evaluation of \eqref{eq:OTOCeikonal} is the bilocal vertex functions. These can easily be constructed by observing that the shocks $X^\pm$ were obtained from performing piecewise $SL(2,\mathbb{R})_{diag}$ transformations along the SK contour. The coupling of the operators to these shocks is induced by performing the same transformations on the vacuum two-point functions. At the full nonlinear level, the $SL(2,\mathbb{R})$ transformations \eqref{eq:FGnonlinear} generating the shocks couple as follows to matter two-point functions:
{\small
\begin{equation}
\begin{split}
 \twopt^\Delta_{42+}(\hat{t}_1,\hat{t}_2) \equiv\frac{1}{2} e^{\Delta\,g_+(\hat{t}_1,\hat{t}_2)} &= \frac{1}{2} \left[ -\frac{1}{{\cal J}^2} \frac{\partial F_{42}(\hat{s}_1) \, \partial \tilde{G}^+_{42}(\hat{s}_2)}{[F_{42}(\hat{s}_1)-\tilde{G}^+_{42}(\hat{s}_2)]^2} \right]^\Delta =\frac{1}{2\left( 1-\frac{1}{2} \sec\left(\frac{\pi v}{2} \right)  \, e^{- \frac{3\pi i}{2}v} \, X^+\right)^{2\Delta}}\\
  \twopt^\Delta_{31-}(\hat{t}_3,\hat{t}_4) \equiv \frac{1}{2}e^{\Delta\,g_-(\hat{t}_3,\hat{t}_4)} &= \frac{1}{2} \left[- \frac{1}{{\cal J}^2} \frac{\partial \tilde{F}^-_{31}(\hat{s}_3) \; \partial G_{31}(\hat{s}_4)}{[\tilde{F}^-_{31}(\hat{s}_3)-G_{31}(\hat{s}_4)]^2} \right]^\Delta=\frac{1}{2\left( 1 + \frac{1}{2} \sec\left( \frac{\pi v}{2} \right) \, e^{\frac{3\pi i}{2} v}\, X^- \right)^{2\Delta}} 
\end{split}
\end{equation}
}\normalsize
where $\Delta$ parametrizes the dimension of the operators. For single-fermion two-point functions, the operator dimension would be $\Delta_\psi=\frac{1}{q}$ (e.g., we have $ \twopt^{\Delta_\psi}_{42+}(\hat{t}_1,\hat{t}_2) = \frac{1}{2}[1+\Delta_\psi^2\,\delta_+g_{42} + \ldots ]$). The non-linear reparametrizations $\tilde{F}_{IJ}^\pm$ and $\tilde{G}_{IJ}^\pm$ were defined in \eqref{eq:FGtildeDef}.
Expanding these bilocal vertex functions in $X^\pm$, we can use the propagator \eqref{eq:Xprop} to evaluate the eikonal integral \eqref{eq:OTOCeikonal}. We find:
\begin{equation}
\begin{split}
\boxed{
   \text{OTOC}_{eik} = \int {dX^+ dX^-\ov {\cal Z}} \; e^{iS_{eik}[X^+,X^-]} \, e^{\Delta\,g_+(\hat{t}_1,\hat{t}_2)}\,e^{\Delta\,g_-(\hat{t}_3,\hat{t}_4)} 
   = z^{-2\Delta} \, U\big(2\Delta,1, z^{-1}\big) }
 \end{split}
 \label{eq:FotocRes}
\end{equation}
where $z=\frac{q^2}{4N}\, \sec \left( \frac{\pi v}{2} \right)^3  e^{-\frac{i\pi v}{2} +vT}$. The leading term in the large $N$ expansion agrees with the expected result \eqref{eq:OTOCres}:
\begin{equation}
\begin{split}
  z^{-2\Delta} \, U\big(2\Delta,1, z^{-1}\big) 
   &= 1-\frac{\Delta^2q^2}{N}\,e^{-\frac{i\pi v}{2}}\,\sec\left(\frac{\pi v}{2} \right)^3 \, e^{vT} + {\cal O}\left(N^{-2}\right) \,. 
 \end{split}
 \label{eq:FotocRes22}
\end{equation}
We found that this OTOC behaves somewhat anomalously for small $\Delta$, e.g. in the fermion four-point function case, where $\Delta_\psi=1/q$. We discuss this in Appendix~\ref{ruelle}.

\subsection{Wave function representation of vertices}

It is again useful to give an integral representation for the bilocal vertices~\cite{Gao:2019nyj,Nezami:2021yaq}, similar to the maximally chaotic cases discussed in section \ref{sec:maximal}:
\beq
e^{\Delta\,g_+(\hat{t}_1,\hat{t}_2)}&=\left( {  \cos \left({\pi v \ov 2}\right)\ov \cosh\left({(i\pi+\hat{t}_1-\hat{t}_2) v\ov 2}\right) -{1\ov 2}e^{-{3i\pi v \ov 2}}e^{-{v(\hat{t}_1+\hat{t}_2)\ov 2}}X^+}\right)^{2\Delta}
\\&=\frac{1}{\Gamma(2\Delta)}{v^{2\Delta} \ov \mathcal J^{2\Delta}} \int_{-\infty}^0 dq_+ {1 \ov - q_+} \Psi_1(q_+) \Psi_2(q_+) \exp\left( -q_+e^{-{3i\pi v \ov 2}}X^+ \right)\,,
\\ e^{\Delta\,g_-(\hat{t}_3,\hat{t}_4)}&=\left( {  \cos \left({\pi v \ov 2}\right)\ov \cosh\left({(i\pi+\hat{t}_3-\hat{t}_4) v\ov 2}\right) +{1\ov 2}e^{{3i\pi v \ov 2}}e^{-{v(2T-\hat{t}_3-\hat{t}_4)\ov 2}}X^-}\right)^{2\Delta}
\\&=\frac{1}{\Gamma(2\Delta)}{v^{2\Delta} \ov \mathcal J^{2\Delta}} \int_{-\infty}^0 dp_- {1 \ov - p_-} \Psi_3(p_-) \Psi_4(p_-) \exp\left(  p_-e^{{3i\pi v \ov 2}}X^- \right)\,,
\eeq
where $\Psi_j\equiv x_j^\Delta e^{-x_j}$ with
\beq
x_1=-q_+ e^{v \left(\hat{t}_1+{i\pi \ov 2}\right)},\, x_2=-q_+ e^{v \left(\hat{t}_2-{i\pi \ov 2}\right)},\, x_3=-p_- e^{v \left(T- \hat{t}_3-{i\pi \ov 2}\right)},\, x_4=-p_- e^{v \left(T- \hat{t}_4+{i\pi \ov 2}\right)}
\eeq
Note that the convergence of the integral is guaranteed since $0<v<1$ and $X^\pm \ll 1$. We therefore do not need to keep track of regulators (c.f.,  \eqref{eq:holoIntegral}).

Using the integral representation, we can write the eikonal OTOC \eqref{eq:FotocRes} as 
{\small
\begin{equation}
\label{eq:FeikonalGrav}
\begin{split}
 \text{OTOC}_{eik} = \frac{1}{\Gamma(2\Delta)^2}\frac{v^{4\Delta}}{{\cal J}^{4\Delta}}\int_{-\infty}^0 dq_+ dp_-  \!\le[ { \Psi_1(q_+) \Psi_2(q_+) \ov -q_+} \ri] \!\le[ { \Psi_3(p_-) \Psi_4(p_-) \ov -p_-} \ri] 
  \exp\left(  \frac{2q^2}{N} \frac{e^{vT-\frac{i\pi v}{2}}}{\cos \left( \frac{\pi v}{2} \right)}\,q_+ p_- \right) 
\end{split}
\end{equation}
}\normalsize
which, of course, evaluates to the same hypergeometric function as in \eqref{eq:FotocRes}. However, this expression makes the connection with gravitational shockwave scattering manifest. 

The authors of~\cite{Nezami:2021yaq} make the intriguing observation that \eqref{eq:FeikonalGrav} through the redefinition $q_+=\mathfrak{q}_+^v,\, p_-=\mathfrak{p}_-^v$ can be rewritten as
{\small
\begin{equation}
\label{eq:FeikonalGrav2}
\begin{split}
 \text{OTOC}_{eik} = \frac{1}{\Gamma(2\Delta)^2}\frac{v^{4\Delta+2}}{{\cal J}^{4\Delta}}\int_{-\infty}^0 d\mathfrak{q}_+ d\mathfrak{p}_-  \!\le[ { \Psi^s_1(\mathfrak{q}_+) \Psi^s_2(\mathfrak{q}_+) \ov -\mathfrak{q}_+} \ri] \!\le[ { \Psi^s_3(\mathfrak{p}_-) \Psi^s_4(\mathfrak{p}_-) \ov -\mathfrak{p}_-} \ri] 
  \exp\left(  \frac{2q^2}{N \cos \left( \frac{\pi v}{2} \right)} \,\le(-ie^{T}\mathfrak{q}_+ \mathfrak{p}_- \ri)^v\right) 
\end{split}
\end{equation}
}\normalsize
where $\Psi^s_i$ are the ``stringy'' wave functions, which differ from the conformal ones defined above through an extra power of $v$ in the $x_j$ arguments:
\begin{equation}
\begin{split}
  \Psi_j^s \equiv  (-{\bf x}_j^v)^{\Delta}\, e^{ {\bf x}_j^v} \quad
  \text{with: } \quad {\bf x}_1 = i \mathfrak{q}_+ e^{\hat{t}_1} \,,\quad {\bf x}_2 = -i\mathfrak{q}_+e^{\hat{t}_2} \,,\quad {\bf x}_3 = -i \mathfrak{p}_- e^{T-\hat{t}_3} \,,\quad {\bf x}_4 = i \mathfrak{p}_- e^{T-\hat{t}_4} \,.
  \end{split}
\end{equation}
We call these wave functions ``stringy'' as they include corrections to the bulk-boundary propagators due to the same effects that give sub-maximal chaos. We are not aware of concrete analogous expressions having been derived in string theory and leave this as an intriguing connection to explore further. On the other hand, the eikonal phase $\le(-ie^{T}\mathfrak{q}_+ \mathfrak{p}_- \ri)^v$ is indeed a known stringy scattering amplitude and can be compared in detail to the corresponding result in \cite{Shenker:2014cwa}; there, the eikonal phase $e^{i\delta_\text{stringy}}$ taking into account tree level string exchange in the scattering of shockwaves on an AdS black hole background is found to be (in our notation):
\begin{equation}
  i\delta_\text{stringy} = \frac{16\pi G_N}{\ell_\text{AdS}^{D-2}} \frac{\ell_\text{AdS}^2}{\ell_s^2} \int \frac{d^{d-1}p}{(2\pi)^{d-1}} \, \frac{e^{ipx}}{p^2 + \mu^2} \left( i \ell_s^2 s / 4 \right)^{1-\ell_s^2(p^2+\mu^2)/(2r_0^2)} \,,\qquad s \sim e^{T} \mathfrak{q}_+ \mathfrak{p}_- 
\end{equation}
where $r_0$ is the horizon radius and $\mu^2 = 2\pi(d-1)r_0/\beta$. Modulo the absence of the momentum integral in our $(0+1)$-dimensional model, this expression is clearly analogous to the phase in \eqref{eq:FeikonalGrav2}. To make the comparison even more precise, we are thus motivated to study a model with spatial extent. This is the subject of the next section.

We note that it was recently argued in \cite{StanfordTalk} that \eqref{eq:FeikonalGrav} can be given a geometric interpretation in terms of the enlarging of the thermal circle to a ``fake thermal circle'' of size $\beta/v$. From that point of view, \eqref{eq:FeikonalGrav} gives the most natural expression. However, the rewriting \eqref{eq:FeikonalGrav2} makes it clear that, at the level of formulas, the mechanism responsible for non-maximal chaos is the same as in string theory. The  ``fake thermal circle'' is likely a non-generic geometric tool to determine the wave functions $\Psi_i$.

\section{The large $q$ SYK chain}
\label{sec:chain}

Let us now consider the SYK chain with sites labelled by $x=0,\ldots M-1$ and periodic boundary conditions in $x$, see \cite{Gu:2016oyy} for details. In addition to the same random Gaussian on-site couplings $J_x$ as before, we now have similar inter-site couplings $J'_x$ between $\frac{q}{2}$ fermions on site $x$ and $\frac{q}{2}$ fermions on site $x+1$. Their respective variances are:
\begin{equation}
\left\langle J_{x,i_1 \cdots i_q}^2 \right\rangle = \frac{(q-1)!}{N^{q-1}} \, J_0^2 \,,\qquad
\left\langle J_{x,i_1 \cdots i_{q/2} j_1 \cdots j_{q/2}}'^{\,2} \right\rangle = \frac{[(q/2)!]^2}{qN^{q-1}} \, J_1^2 \,,\qquad  {\cal J}_{0,1}^2 \equiv \frac{q}{2^{q-1}} \,J_{0,1}^2\,.
\end{equation}
In the large $q$ limit, we hold ${\cal J}_{0,1}^2$ fixed. We will also use ${\cal J}^2 \equiv {\cal J}_0^2 + {\cal J}_1^2$ and $\gamma \equiv {\cal J}_1^2/{\cal J}^2$.
The effective action for the chain on the same generalized Schwinger-Keldysh contour as before, takes the form of a lattice of Liouville theories with nearest-neighbor couplings:
\beq
\label{eq:Schain}
iS={N\ov 4q^2} \sum_{x=0}^{M-1}  \int ds_1 ds_2 \left[  {1\ov 4}\pa_1 g_{{\scriptscriptstyle IJ},x} \,\pa_2 g_{{\scriptscriptstyle IJ},x} -\mathcal{J}_0^2  r_{s_1} r_{s_2}\,e^{g_{\scaleto{IJ}{3.5pt},x}} -\mathcal{J}_1^2  r_{s_1} r_{s_2}\,e^{\frac{1}{2}(g_{\scaleto{IJ}{3.5pt},x}+g_{\scaleto{IJ}{3.5pt},x+1})} \right]\,,
\eeq
Note that the fermion bilocals curiously only have one spatial argument, reflecting the fact that only groups of $q/2$ fermions can hop between lattice sites.

The leading connected contribution to the OTOC between two fermions each on sites $x$ and $x'$ was computed analytically in the large $q$ limit in \cite{Choi:2020tdj}, see also \cite{Gu:2018jsv}. The goal of this section is to reproduce these results using the eikonal action derived from the path integral formulation, and generalize it to higher orders. Due to the spatial extent, we will see novel features compared to the single SYK site.

\subsection{Eikonal action for the large $q$ SYK chain}

The equation of motion for $g_{{\scriptscriptstyle IJ},x}(s_1,s_2)$ derived from \eqref{eq:Schain} is:
\begin{equation}
\label{eq:eomChain}
\frac{1}{2} \partial_1\partial_2 g_{{\scriptscriptstyle IJ},x} - {\cal J}^2 r_{s_1} r_{s_2}  \left[ (1-\gamma) \, e^{g_{\scaleto{IJ}{3.5pt},x}} + \frac{\gamma}{2} \left( e^{\frac{1}{2} (g_{\scaleto{IJ}{3.5pt},x}+g_{\scaleto{IJ}{3.5pt},x+1})} + e^{\frac{1}{2}( g_{\scaleto{IJ}{3.5pt},x-1}+g_{\scaleto{IJ}{3.5pt},x})} \right) \right]=0\,.
\end{equation}
A particular $x$-independent saddle point solution $g_{IJ}$ exists, which is of the same form as in previous sections:
\beq
e^{g_{IJ}(s_1,s_2)} ={1\ov \mathcal J^2\, r_{s_1} r_{s_2}}{F_{IJ}'(s_1) G_{IJ}'(s_2)\ov (F_{IJ}(s_1)-G_{IJ}(s_2))^2}\qquad (s_1\in I, s_2\in J)\,.
\eeq
Let us now consider fluctuations around this saddle point. These will a priori be distinct for every site $x$. The quadratic action is 
\begin{equation}
\begin{split}
  iS_{quad} &= \frac{N}{4q^2} \sum_{x,y} \big\langle \delta g_x \,,\, {\cal L}_{x,y} \, \delta g_y \big\rangle  \\
  {\cal L}_{x,y} &\equiv   -\delta_{x,y} \left[ \frac{1}{4} \, \partial_1 \partial_2 + \left( \frac{{\cal J}_0^2}{2} + \frac{{\cal J}_1^2}{4} \right)  r_{s_1} r_{s_2} \, e^{g_{IJ}} \right] - \frac{{\cal J}_1^2}{4} \, r_{s_1} r_{s_2} \, e^{g_{IJ}}\, (\delta_{x,y-1} + \delta_{x,y+1})
 \end{split}
\end{equation}
From the translational invariance of the large $q$ SYK chain, we can analyze the eikonal action for each momentum $p$. After a Fourier transform, we get
\beq
iS_{quad} &={N \ov 4q^2}\int_{-\pi}^{\pi} \frac{dp}{2\pi} \ex{ \del g_{-p} ,\,\mathcal L_p\, \del g_p}
\\ \mathcal L_p &\equiv - \left[{1\ov 4}\pa_1\pa_2 +\left( \frac{{\cal J}_0^2}{2} + \frac{{\cal J}_1^2(1+\cos (p))}{4} \right)  r_{s_1} r_{s_2} \, e^{g_{IJ}}\right] \\
&= - \left[{1\ov 4}\pa_1\pa_2 + \frac{{\cal J}^2}{2} \, \frac{h(h-1)}{2}\, r_{s_1} r_{s_2} \, e^{g_{IJ}}\right]
\eeq
Similary, we turn on zero modes for each momentum $p$ of the form
\beq
~& \del g_{{\scriptscriptstyle IJ},\,p}(t_1,t_2)=X^+(p)\,\del_+ g_{{\scriptscriptstyle IJ},p}(t_1,t_2)+X^-(p)\,\del_- g_{{\scriptscriptstyle IJ},p}(t_1,t_2)\,.
\eeq
The near-zero modes satisfying KMS conditions are given by
\beq
\label{eq:deltagCHAIN}
~&\de_\pm g_{{\scriptscriptstyle IJ},\,p}(t_1,t_2)=A_{\pm,{\scriptscriptstyle IJ}}(p)\le[e^{\pm v ((1\mp 1)T-t_1-t_2)/2}\ov \cos\le({v\ov2}\le(\pi-it_{12}\ri)\ri)\ri]^{h(p)-1} \,,\qquad \frac{h(h-1)}{2} = 1 + \frac{\gamma}{2} \, [\cos(p) - 1] \,, 
\eeq
where $\gamma \equiv {\cal J}_1^2/{\cal J}^2$ and $A_{\pm, IJ}$ for $I\geq J$ are given by 
\beq
\\&A_{+,IJ}(p) =\begin{cases}e^{-\frac{i\pi \kappa}{2} } &\quad(I,J)=(2,1),(3,1)
\\
e^{-\frac{3i\pi \kappa}{2} } &\quad(I,J)=(4,2),(5,2),(4,3),(5,3)
\\0 &\quad 	\text{otherwise}
\end{cases}
\\& A_{-,IJ}(p)=\begin{cases}-e^{\frac{3i\pi \kappa}{2} } &\quad(I,J)=(3,1),(4,1),(3,2),(4,2)
\\
-e^{\frac{5i\pi \kappa}{2}}&\quad(I,J)=(5,3),(5,4)
\\0 &\quad 	\text{otherwise}
\end{cases}
\eeq
where $\kappa \equiv v(h-1)$ is the Lyapunov exponent.
As before, these coefficients are only determined up to an overall rescaling, which drops out in the calculation of correlation functions. Following the strategy of section \ref{sec:actionCalc}, we evaluate the action $S_{quad}$ along the out-of-time-order contour on these near-zero modes, and we reduce the computation to boundary terms in the time domain (for fixed spatial momentum $p$).
After some rather non-trivial algebraic simplifications, we find that the eikonal action thus obtained can be written as:
\beq \label{EikonalChain}
\boxed{
\begin{split}
iS_{eik} &=  \int_{-\pi}^\pi \frac{dp}{2\pi} \; iC_{eik}(p)  \,X^+(p)X^-(-p)\,,\\ 
iC_{eik}(p) &= -\frac{N}{2q^2}{e^{\frac{i\kappa \pi}{2}-\kappa T} \ov  \sqrt{\pi}} \cos\left( \frac{\pi \kappa}{2} \right)  \cos \left( \frac{\pi \kappa}{v} \right) \Gamma \left( \frac{\kappa}{v}+1 \right) \Gamma \left(\frac{1}{2} - \frac{\kappa}{v} \right) 
\end{split}
}
\eeq 
For $p=0$, we have $h(p=0)=2$ and $\kappa(p=0)=v$, so the action reduces to
\beq
iC_{eik}(p=0)={N\ov q^2}\cos\left({\pi v\ov 2}\right) e^{{i\pi v \ov 2}-vT} \,.
\eeq
This reproduces the eikonal action of the isolated SYK `dot', \eqref{eq:Sdot}.

\subsection{Leading connected OTOC from eikonal action}

We now have all the ingredients to compute the leading connected contribution to the OTOC from a single propagator $\langle X^+(p) X^-(-p)\rangle = \left(iC_{eik}(p)\right)^{-1}$:
\begin{equation}
\label{eq:OTOCchainRes}
\begin{split}
\frac{1}{N} {\cal F}^\text{otoc}(\hat{t}_1,\hat{t}_2,\hat{t}_3,\hat{t}_4;x) 
 &= \int \frac{dp}{2\pi} \, e^{ipx} \, \big( \Delta\,\delta_+ g_{42,p}(\hat{t}_1,\hat{t}_2) \big)\,\big( \Delta\,\delta_- g_{31,-p}(\hat{t}_3,\hat{t}_4) \big)\,\left\langle X^+(p) X^-(-p) \right\rangle \\
 &= -\frac{2\sqrt{\pi}}{N}\int \frac{dp}{2\pi} \, \frac{e^{ipx + \kappa\left(-\frac{i\pi}{2}+T\right)}}{\cos \big( \frac{\pi \kappa}{2} \big) \cos \big( \frac{\pi \kappa}{v} \big) \Gamma\big( \frac{\kappa}{v}+1 \big) \Gamma\big( \frac{1}{2} - \frac{\kappa}{v}\big)} \left[ \frac{1}{\cos \big( \frac{\pi v}{2} \big) }\right]^{{2\kappa \ov v}}
\end{split}
\end{equation}
where we chose the spatial insertions at $x$ and $x'=0$ without loss of generality. This result permits a stringent check on $C_{eik}(p)$: in \cite{Gu:2018jsv} a technology was developed to compute the leading order in $1/N$ OTOC for theories dominated by ladder diagrams. We use their technology in Appendix~\ref{GKcompare} to show that \eqref{eq:OTOCchainRes} matches the diagrammatic results.

To do the momentum integral, one has to distinguish two contributions (c.f.\ \cite{Choi:2020tdj,Mezei:2019dfv}):
\begin{enumerate}
  \item {\bf Saddle point:} When $u\equiv x/T$ is small, then the saddle point momentum $p_*$ extremizing $ipx+\kappa\big(-\frac{i\pi}{2}+T \big)$ dominates the integral. It is given by:
  \begin{equation}
   p_*  =  i\, \text{arccosh} \left[  \tilde{u}^2 + \sqrt{  \tilde{u}^4 + \frac{1}{{2\gamma}}\left( 9-4\gamma\right)\tilde{u}^2 + 1 }\, \right] \in i \, \mathbb{R} \,,\qquad \tilde{u} \equiv  \frac{x}{\left(T -\frac{i\pi}{2} \right)}\sqrt{\frac{2}{\gamma v^2}}\,.
   \label{eq:pstar}
  \end{equation}
  For small $u$, this is $p_* \approx  \frac{3i}{\sqrt{2\gamma}} \, \tilde{u} + {\cal O}(\tilde{u}^2)$, so the integration contour capturing the saddle point gets deformed in the imaginary direction. The $1/N$ piece of the OTOC takes the following form for small $\tilde u$ as a function of $T$ and $x$:
    \begin{equation}
    \label{eq:cfStringy}
  \begin{split}
     \frac{1}{N}{\cal F}^\text{otoc} \Big{|}_\text{saddle} 
     &= \frac{c(v,\gamma)}{N}\times e^{v\left(-\frac{i\pi}{2}+T\right)} \times\frac{e^{- \frac{x^2}{2r^2}}}{\sqrt{2\pi r^2}}  \,,\qquad r^2 = \frac{\gamma v}{3}\left( - \frac{i\pi}{2}+T \right)
   \end{split}
  \end{equation}
  where $c(v,\gamma)$ is a prefactor that can be worked out easily from \eqref{eq:OTOCchainRes}. We deliberately wrote this expression in a similar form as Eq.\ (70) of \cite{Shenker:2014cwa} in order to facilitate a comparison with effects of string scattering in gravity: the second factor above comes from evaluating $e^{\kappa(h-1)(-\frac{i\pi}{2}+T)}$ on the saddle point $p_*$ for small $x$; it exhibits exponential growth of the form $e^{vT}$ as in the SYK dot, i.e., with Lyapunov exponent $\lambda_L = v$. The spatial dependence (third factor) comes from evaluating $e^{ip_* x}$ and performing the Gaussian integral after expanding the exponent in \eqref{eq:OTOCchainRes} around the saddle point to second order; it consists of a `saddle point value' $e^{-x^2/(2r^2)}$ and a `Gaussian determinant' $1/\sqrt{2\pi r^2}$.

Alternatively, we can also fix the velocity $u=x/T$ and consider the Lyapunov exponent as a function of $u$, see \cite{Mezei:2019dfv}. This {\it velocity dependent Lyapunov exponent} takes the form 
\begin{equation}
\lambda_L(u) = v \left( 1- \frac{3}{2\gamma v^2} \, u^2 + {\cal O}(u^4) \right)
\end{equation}
  \item {\bf Pole at $\kappa=1$:} When $x/T$ is sufficiently large, the pole at $\kappa=1$ due to $\cos \big(\frac{\pi\kappa}{2}\big)$ in the denominator dominates the integral. Its contribution is:
  \begin{equation}
  \begin{split}
     \frac{1}{N} {\cal F}^\text{otoc} \Big{|}_\text{pole} 
     &= \frac{4v}{N\sqrt{\pi}} \frac{e^{T - x/u_B^{(T)} }}{\cos \left( \frac{\pi}{v} \right) \Gamma\left(\frac{1}{v} \right) \Gamma \left( \frac{1}{2} - \frac{1}{v} \right) \cos \left( \frac{\pi v}{2} \right)^{\frac{2}{v}}} \,,\\
      \text{with:} \;\;\; u_B^{(T)} &=  \left[\text{arccosh} \left( \frac{1+v+(\gamma-2)v^2}{\gamma v^2} \right) \right]^{-1}\,,
    \end{split}
  \end{equation}
  where $u_B^{(T)}$ is the `stress tensor' associated butterfly velocity. To leading order as $v\rightarrow 1$, the result is ${\cal F}^\text{otoc}|_\text{pole} \approx \frac{8}{\pi^3(1-v)^2} \, e^{T-x/u_B^{(T)}}$ with $1/u_B^{(T)} \approx  \sqrt{\frac{6}{\gamma}(1-v)}$. In other words, the velocity dependent Lyapunov exponent takes the maximal value $\lambda_L(u) = 1 - u/u_B^{(T)}$.
  It is useful to note the momentum location of the pole, which is given by:
  \begin{equation}
    p_\text{pole}=  \frac{i}{u_B^{(T)}}= \text{arccos} \left( \frac{1+v+(\gamma-2)v^2}{\gamma v^2} \right) \in i\,\mathbb{R}_+
  \end{equation}
  We can see that the pole becomes relevant -- and in fact dominant -- when the saddle point integration contour through $p_*$ crosses $p_\text{pole}$, or equivalently when
  \begin{equation}
  u\equiv \frac{x}{T} \;\longrightarrow\; {u}_* = \frac{1}{v+2} \sqrt{\left( 1+v-2v^2\right)\left( 1+v+2(\gamma-1)v^2\right)} 
  \end{equation}
  We refer to \cite{Choi:2020tdj} for a more detailed discussion of this mechanism.
\end{enumerate}

\subsection{Vertices of the SYK chain}

In the SYK chain the near zero modes take the form \eqref{eq:deltagCHAIN}, i.e., they are just the zero modes of the SYK `dot' raised to the power $(h(p)-1)$. In particular, we have
\begin{equation}
  \delta_+ g_{42,p}(\hat{t}_1,\hat{t}_2) = e^{-\frac{3i\pi\kappa}{2}} \left[ \frac{1}{\cos \big(\frac{\pi v}{2}\big)} \right]^{h(p)-1} \,,\quad \delta_- g_{31,p}(\hat{t}_3,\hat{t}_4) = -e^{\frac{3i\pi\kappa}{2}} \left[ \frac{1}{\cos \big(\frac{\pi v}{2}\big)} \right]^{h(p)-1}\,.
\end{equation}
We would like to generalize these to nonlinear vertex functions $e^{\Delta \, g_{IJ,p}^\pm(t,t')}$ sourced by $X^\pm(p)$.
We draw inspiration from the linearized soft modes to guess the vertex factor to be:
\es{VertexFactor}{
e^{\Delta g_{42,x}^+(t_1,t_2)}&=\frac{1}{\Gamma(2\Delta)}{v^{2\Delta} \ov \mathcal J^{2\Delta}} \int_{-\infty}^0 dq_+\ {1 \ov - q_+} \Psi_1(q_+) \Psi_2(q_+) \\
 &\qquad\qquad\;\; \times \exp\left\{ \int \frac{dp}{2\pi} \frac{ \Gamma(2\Delta+1)\,2^{h(p)-2}}{\Gamma(2\Delta+h(p)-1)} \  (-q_+)^{\color{red}{h(p)-1}} e^{-{3i\pi \kappa(p) \ov 2}}X^+(p)\, e^{ipx} + \ldots \right\}\\
&=e^{\Delta g_{42}(t_1,t_2)}  \le[1+  \int \frac{dp}{2\pi} \,\Delta\,\de_+ g_{42,p}(t_1,t_2)\,X^+(p)\, e^{i p x} + \ldots  \ri]
}
where $g_{IJ}$ is the $x$-independent saddle point and we highlighted the crucial difference compared to the single-site SYK model. In the exponent, we chose a normalization such that our previous conventions are reproduced (recall that in momentum space, we had $e^{\Delta g^\pm_{IJ,p}} =e^{\Delta g_{IJ}}  (1 + \Delta \, \delta_\pm g_{IJ,p} \, X^\pm(p)+ \ldots)$). Note also that $\frac{ \Gamma(2\Delta+1)\,2^{h(p)-2}}{\Gamma(2\Delta+h(p)-1)} \rightarrow 1$ as $h(p) \rightarrow 2$ such that the integral representation for the large $q$ SYK dot is reproduced at $p=0$. Similarly:
\es{VertexFactorM}{
e^{\Delta g^-_{31,x'}( t_3,t_4)}&=\frac{1}{\Gamma(2\Delta)}{v^{2\Delta} \ov \mathcal J^{2\Delta}} \int_{-\infty}^0 dp_-\ {1 \ov - p_-} \Psi_3(p_-) \Psi_4(p_-) \\
 &\qquad\; \times \exp\left\{ -\int \frac{dp'}{2\pi}  \  \frac{ \Gamma(2\Delta+1)\,2^{h(p)-2}}{\Gamma(2\Delta+h(p)-1)} \, (-p_-)^{\color{red}{h(p')-1}} e^{{3i\pi \kappa(p') \ov 2}}X^-(p')\, e^{ip'x'} + \ldots \right\}
}

Unlike in the $(0+1)$-dimensional (or $p=0$) case, the above vertex factors do not solve the equation of motion \eqref{eq:eomChain}.
Instead, we should think of the solution as being constructed order by order in $X^\pm(p)$. At higher orders in this expansion the solution takes the following form: 
{\small
\es{VertexFactor2}{
e^{\Delta g^+_{42,x}(t_1,t_2)}&=\frac{1}{\Gamma(2\Delta)}{v^{2\Delta} \ov \mathcal J^{2\Delta}} \int_{-\infty}^0 dq_+\ {1 \ov - q_+} \Psi_1(q_+) \Psi_2(q_+) \\
&\quad\;\; \times \exp\bigg\{ \int \frac{dp_1}{2\pi}  \, b^+_1\, X^+_1 \\
 &\qquad\qquad\;\; +  \iint \frac{dp_1}{2\pi}\frac{dp_2}{2\pi}   \left(A_2(p_1,p_2) - \frac{1}{2} \right) \,  b^+_1 b^+_2\, X^+_1 X^+_2 \\
 &\qquad\qquad\;\; +  \iiint \frac{dp_1}{2\pi}\frac{dp_2}{2\pi}\frac{dp_3}{2\pi}  \big( A_3(p_1,p_2,p_3) - \ldots \big) \, b^+_1 b^+_2 b^+_3 \, X^+_1 X^+_2 X^+_3  + \ldots \bigg\}
}
}\normalsize
where $b^+_k \equiv b^+(p_k) =  \frac{ \Gamma(2\Delta+1)\,2^{h(p_k)-2}}{\Gamma(2\Delta+h(p_k)-1)} \,(-q_+)^{h(p_k)-1} \, e^{-\frac{3i\pi \kappa(p_k)}{2}} \, e^{ip_kx}$. This yields, order by order in $X^+$:
{\small
\es{VertexFactor2b}{
&e^{\Delta g^+_{42,x}(t_1,t_2)}= e^{\Delta g_{42}(t_1,t_2)} \bigg\{ 1 
+  \Delta\int \frac{dp_1}{2\pi}\, \left[ \delta_+ g_{42}(t_1,t_2) \right]^{h(p_1)-1} X^+(p_1) \, e^{ip_1 x}\\
&\qquad\qquad\qquad\qquad\;\, +\Delta \int \frac{dp_1dp_2}{(2\pi)^2}\,\frac{ \mathfrak{b}_{\Delta,p_1,p_2}}{\mathfrak{b}_{\Delta,p_1}\,\mathfrak{b}_{\Delta,p_2}}\, A_2(p_1,p_2) \, \left[ \delta_+ g_{42}(t_1,t_2) \right]^{h(p_1)+h(p_2)-2} X^+(p_1) X^+(p_2) \, e^{i(p_1+p_2)x} \\
&\qquad\qquad\qquad\qquad\;\,+  \ldots \bigg\} 
}
}\normalsize
where $\delta_\pm g_{IJ}(t,t')$ are the zero-momentum expressions as in \eqref{eq:shocksimple} and we defined 
\begin{equation}
\mathfrak{b}_{\Delta,p_1,\ldots,p_n}  \equiv \frac{\Gamma(2\Delta+h(p_1)+\ldots + h(p_n) - n)}{\Gamma(2\Delta+1)2^{h(p_1)+\ldots + h(p_n) - n-1}}\,.
\end{equation}
The pattern continues as indicated at higher orders, and the integral kernels encode the degree to which the vertex factor fails to factorize into powers of a single momentum integral.

Requiring that $1/\Delta$ times the logarithm of the expression \eqref{VertexFactor2b} solves the $g_x$ equation of motion \eqref{eq:eomChain} determines the functions $A_n(p_1,\ldots,p_n)$:
{\small
\es{A2expr}{
A_2&= \frac{\mathfrak{b}_{\Delta,p_1} \, \mathfrak{b}_{\Delta,p_2}}{\mathfrak{b}_{\Delta,p_1,p_2}} \left[ \frac{2(h_1+h_2-1)(h_1+h_2-2)+\gamma ( \cos(p_1) + \cos(p_2) - \cos(p_1+p_2)-1 )}{4(h_1+h_2)(h_1+h_2-3) + 4\gamma (1-\cos(p_1+p_2))} - \frac{1-\Delta}{2} \right] \\
 A_3&=  \frac{\mathfrak{b}_{\Delta,p_1} \, \mathfrak{b}_{\Delta,p_2}\, \mathfrak{b}_{\Delta,p_3}}{\mathfrak{b}_{\Delta,p_1,p_2,p_3}} \times \dots\\
& \;\;\vdots
}
}\normalsize
where $h_i \equiv h(p_i)$.
The first coefficient, $A_1(p)$, is not fixed by the equations of motion. It corresponds to a normalization of $X^+(p)$, which we previously fixed.

To summarize, we get the following expansion for the vertex factor in powers of $X^+(p)$:
{\small
\begin{equation}
\label{eq:vertexChainFull}
\begin{split}
 &e^{\Delta \, g^+_{42,x}(t_1,t_2)} = e^{\Delta\,g_{42}(t_1,t_2)}\Bigg\{1+\Delta \sum_{n\geq 1}  \int \frac{dp_1 \cdots dp_n}{(2\pi)^n} \,\frac{\mathfrak{b}_{\Delta,p_1,\ldots,p_n}}{\mathfrak{b}_{\Delta,p_1}\cdots \mathfrak{b}_{\Delta,p_n}}\, A_n(p_1,\ldots,p_n)  \\
  &\qquad\qquad\qquad\qquad\qquad\qquad\qquad\quad \times  \left[ e^{-\frac{3i\pi v}{2}}\, \frac{e^{-v(t_1+t_2)/2}}{\cos \left( \frac{v}{2} (\pi - i t_{12})\right)} \right]^{h_1 + \ldots + h_n - n}\,X^+_1 \cdots X^+_n \;e^{i(p_1 + \ldots + p_n)x}  \Bigg\}
\end{split}
\end{equation}
}\normalsize
and similarly for $e^{\Delta\,g_{31,x}^-(t_1,t_2)}$.

Note that the coefficients in the vertex factor expansion \eqref{VertexFactor2} have a nice structure: even though we derived them in an amplitude expansion (in powers of $X^\pm$), one can easily verify that they simultaneously provide a momentum expansion. For instance:
\begin{equation}
  \begin{split}
   A_2(p_1,p_2) - \frac{1}{2} &= -\frac{\gamma}{4+8\Delta} \, p_1p_2 + {\cal O}(p_i^4) \,. \label{softlim}
  \end{split}
\end{equation} 
Higher order terms $A_{n\geq 3}$ (in the combination that appears in the exponent of \eqref{VertexFactor2}) scale with successively higher powers of the momenta in this `soft limit'.\footnote{ Note that these higher order terms are {\it not} suppressed compared to $A_2$ in the limit of small $\Delta$.}

\subsection{Path integral with higher orders in the shock amplitudes}

Using the vertex function \eqref{eq:vertexChainFull}, we can in principle compute the OTOC in the large $q$ SYK chain order by order in the $1/N$ expansion. We get the following structure:
\begin{equation}
\begin{split}
 \text{OTOC}_{eik} &= \int\frac{ dX^+(p) dX^-(p)}{{\cal Z}(p)} \, e^{iS_{eik}[X^+,X^-]} \, e^{\Delta \, g^+_{42,x}(\hat{t}_1,\hat{t}_2)} \, e^{\Delta\, g^-_{31,x'}(\hat{t}_3,\hat{t}_4)} \\
 &= 1 + \Delta^2 \sum_{n\geq 1} \int  \frac{dp_1 \cdots dp_n}{(2\pi)^n} \,\left[ \frac{\mathfrak{b}_{\Delta,p_1,\ldots,p_n}}{\mathfrak{b}_{\Delta,p_1}\cdots \mathfrak{b}_{\Delta,p_n}}\, A_n(p_1,\ldots,p_n) \right]^2 \\
  &\qquad\qquad\qquad\qquad \times  \left[\frac{1}{\cos \left( \frac{\pi v}{2} \right)} \right]^{2(h_1 + \ldots + h_n - n)} \frac{n!}{iC_{eik}(p_1) \cdots iC_{eik}(p_n)} \, e^{i(p_1+\ldots+p_n)(x-x')}
\end{split}
\end{equation}
where $C_{eik} \sim \Delta^2N\, e^{-vT}$ contains the time dependence.
This expression can be analyzed in terms of pole and saddle-point contributions, as we did for the first term at ${\cal O}\big(\frac{1}{N}\big)$. However, we will find it more illuminating to give a different representation, which resembles the physics of gravitational scattering more directly.

Consider the following integral, where we write both vertex functions in their wave function representation and explicitly include terms quadratic in the shock amplitude (and omit for simplicity any terms at cubic or higher order):
\begin{equation}
\begin{split}
\text{OTOC}_{eik}&= \frac{1}{\Gamma(2\Delta)^2}\frac{v^{4\Delta}}{{\cal J}^{4\Delta}}  \int\frac{ dX^+(p) dX^-(p)}{{\cal Z}(p)} \int_{-\infty}^0 dq_+ dp_-\!  \left[ \frac{\Psi_1(q_+)\Psi_2(q_+)}{-q_+} \right]\! \left[ \frac{\Psi_3(q_-)\Psi_4(q_-)}{-q_-} \right] \\ 
 & \quad \times \text{exp} \bigg\{  \int \frac{dp}{2\pi} \, \vec{X}(p)^\text{T} \cdot \vec{b}(p) -  \frac{1}{2} \int \frac{dp_1}{2\pi} \frac{dp_2}{2\pi} \; \vec{X}(p_1)^\text{T} \cdot {\bf M}(p_1,p_2) \cdot \vec{X}(p_2) + \ldots \bigg\}
 \end{split}
 \label{eq:Fcalc9}
\end{equation}
where 
\begin{equation}
\begin{split}
 \vec{X}(p) &\equiv \begin{pmatrix} X^+(p) \\ X^-(p)  \end{pmatrix}  \,,\quad
 \vec{b}(p) \equiv  \begin{pmatrix} b^+(p) \\ b^-(p) \end{pmatrix} \equiv \begin{pmatrix}  \frac{1}{\mathfrak{b}_{\Delta,p}}\,(-q_+)^{h(p)-1} e^{-\frac{3i\pi \kappa}{2}} e^{ipx}  \\  - \frac{1}{\mathfrak{b}_{\Delta,p}} \,(-p_-)^{h(p)-1} e^{\frac{3i\pi \kappa}{2}} e^{-ipx'}\end{pmatrix} \\
{\bf M}(p_1,p_2) &\equiv \begin{pmatrix} \left(1-2A_2(p_1,p_2)\right) b^+(p_1)b^+(p_2) &-  iC_{eik}(p_1)\,2\pi  \delta(p_1-p_2) \\ - iC_{eik}(p_1) \,2\pi \delta(p_1-p_2) & \left(1-2A_2(p_1,p_2)\right)b^-(p_1)b^-(p_2)\end{pmatrix}  \,.
 \end{split}
\end{equation}
Performing this Gaussian integral, we find for the eikonal integral:
{\small
\begin{equation}
\begin{split}
\text{OTOC}_{eik} &=  \frac{1}{\Gamma(2\Delta)^2}\frac{v^{4\Delta}}{{\cal J}^{4\Delta}} \int_{-\infty}^0 dq_+ dp_- \, \left[ \frac{\Psi_1(q_+)\Psi_2(q_+)}{-q_+} \right] \left[ \frac{\Psi_3(p_-)\Psi_4(p_-)}{-p_-} \right] \\
&\quad \times \exp \bigg\{  \int \frac{dp}{2\pi}\, \frac{1}{\mathfrak{b}_{\Delta,p}^2} \frac{(q_+p_-)^{h(p)-1}}{iC_{eik}(p)} \, e^{ip(x-x')}   \\
 &\qquad\;\;
  + \int \frac{dp_1}{2\pi} \frac{dp_2}{2\pi}  \, \frac{1}{\mathfrak{b}_{\Delta,p_1}^2\mathfrak{b}_{\Delta,p_2}^2}\left( 2A_2^2 - \frac{1}{2}  \right)\frac{(q_+p_-)^{h(p_1)+h(p_2)-2}}{iC_{eik}(p_1)\,iC_{eik}(p_2)} \, e^{i(p_1+p_2)(x-x')}+ \ldots 
   \bigg\}
 \end{split}
 \label{eq:SeikFinal}
\end{equation}
}\normalsize
which is organized in powers of $N$.\footnote{One can further evaluate \eqref{eq:SeikFinal} if desired, using the following general identity:
\begin{equation}
\begin{split}
& \frac{v^{4\Delta}}{{\cal J}^{4\Delta}}\int_{-\infty}^0 dq_+ dp_- \!\left[ \frac{\Psi_1(q_+)\Psi_2(q_+)}{-q_+} \right] \left[ \frac{\Psi_3(p_-)\Psi_4(p_-)}{-p_-} \right] (q_+ p_-)^\alpha = \\
&\qquad =\frac{v^{4\Delta}}{{\cal J}^{4\Delta}} \frac{ \Gamma(\alpha+2\Delta)^2\, e^{-\frac{\alpha v}{2}(2T+\hat{t}_1+\hat{t}_2-\hat{t}_3-\hat{t}_4)}}{\left[ 4\cosh\left(\frac{v}{2}(i\pi + \hat{t}_{12})\right) \cosh\left(\frac{v}{2}(i\pi + \hat{t}_{34})\right) \right]^{\alpha+2\Delta}}
=\frac{\Gamma(\alpha+2\Delta)^2}{\left[2\cos \left(\frac{\pi v}{2} \right)\right]^{2\alpha} }\,,
 \end{split}
\end{equation}
where we inserted our OTOC configuration in the last step.} Finally, following the observation in \cite{Nezami:2021yaq}, we can redefine $q_+,\, p_-$ as we did in \eqref{eq:FeikonalGrav2} and obtain:
{\small
\begin{equation}
\boxed{
\begin{split}
\text{OTOC}_{eik} &=  \frac{1}{\Gamma(2\Delta)^2}\frac{v^{4\Delta+2}}{{\cal J}^{4\Delta}} \int_{-\infty}^0 d\mathfrak{q}_+ d\mathfrak{p}_- \, \left[ \frac{\Psi^s_1(\mathfrak{q}_+)\Psi^s_2(\mathfrak{q}_+)}{-\mathfrak{q}_+} \right] \left[ \frac{\Psi^s_3(\mathfrak{p}_-)\Psi^s_4(\mathfrak{p}_-)}{-\mathfrak{p}_-} \right] \\
&\quad \times \exp \bigg\{  \int \frac{dp}{2\pi}\, \frac{1}{\mathfrak{b}_{\Delta,p}^2} \frac{(-i\mathfrak{q}_+\mathfrak{p}_- e^T)^{\kappa(p)}}{i\mathfrak{C}(p)} \, e^{ip(x-x')}   \\
 &\qquad\;\;
  + \int \frac{dp_1}{2\pi} \frac{dp_2}{2\pi}  \, \frac{1}{\mathfrak{b}_{\Delta,p_1}^2\mathfrak{b}_{\Delta,p_2}^2}\left( 2A_2^2 - \frac{1}{2}  \right)\frac{(-i\mathfrak{q}_+ \mathfrak{p}_- e^T)^{\kappa(p_1)+\kappa(p_2)}}{i\mathfrak{C}(p_1)\,i\mathfrak{C}(p_2)} \, e^{i(p_1+p_2)(x-x')}+ \ldots 
   \bigg\}
 \end{split}
 }
  \label{eq:SeikFinal2}
\end{equation}
}\normalsize
where we pulled out the time dependence and the nontrivial phase factor from $C_{eik}(p)$ to get
\es{Creduced}{
i\mathfrak{C}(p)= -\frac{N}{2q^2 \sqrt{\pi}} \cos\left( \frac{\pi \kappa}{2} \right)  \cos \left( \frac{\pi \kappa}{v} \right) \Gamma \left( \frac{\kappa}{v}+1 \right) \Gamma \left(\frac{1}{2} - \frac{\kappa}{v} \right) \,.
}

The structure of the expression \eqref{eq:SeikFinal2} is rather appealing, as it can be given a string scattering interpretation in complete analogy with the stringy effects in holographic gauge theories \cite{Shenker:2014cwa}. The Mandelstam $s$ invariant in the scattering process is $s=-i\mathfrak{q}_+ \mathfrak{p}_- e^T$, which gets raised to the power $J-1$, where $J$ is the spin of the string's Regge trajectory parametrized by the transverse momentum $p$, with  $J=1+\kappa(p)<2$. Finally, $\mathfrak{C}(p)$ plays the role of $1/G_N$ and the $\Psi^c_i$ are to be thought of as bulk to boundary propagators, converting the microscopic operators into some collective degrees of freedom that can exchange strings. If we only had $A_2=1/2$, the formula would have an interpretation in terms of the eikonalization of single string exchanges. The second line is suggestive of multi-string exchanges, which in the SYK chain is not suppressed at intermediate values of the coupling $v$. However, in the strongly coupled limit $v\to 1$ these integrals, and more generally all physical processes are dominated by $p^2=O(1-v)$ \cite{Gu:2016oyy,Choi:2020tdj} and the nice soft behavior of $A_2$ given in \eqref{softlim} makes these effects negligible. It would be fascinating to understand the analogous multi-string effects in AdS/CFT.

The same remarks apply as around \eqref{eq:FeikonalGrav2}: the rewriting of \eqref{eq:SeikFinal} in \eqref{eq:SeikFinal2} makes it clear that the mechanism for non-maximal chaos in the SYK chain is the same as in string theory. A factor of $v$ from $\kappa(p)$ and the wave functions $\Psi_i$ could be explained from the ``fake thermal circle'' interpretation of \cite{StanfordTalk}, but this is insufficient to explain the $p$-dependence of $\kappa(p)$. This is in line with the speculation put forward in \cite{StanfordTalk} that non-maximal SYK scrambling may come from a sum of a ``fake thermal circle'' and an inherently stringy non-geometric component.

\section{Discussion}

In this paper we have shown in several examples how to derive an effective description of quantum chaos. For maximally chaotic systems (the Schwarzian theory and the Virasoro identity block sector of two-dimensional CFTs) we generalized previous work to give a streamlined derivation of the eikonal action. To study sub-maximal chaos, we considered the large $q$ SYK model and its spatially extended generalization. We will now give some comments on our findings and on future directions. In all cases it was possible to derive the eikonal action by analyzing nearly-zero modes on the double Keldysh time contour. These give an enhanced contribution to the path integral at large time separation, and we showed how to perform the path integral over these modes exactly.

In the case of the Schwarzian or the Alekseev-Shatshvili action the degrees of freedom are reparametrization modes and the $X^\pm$ scramblons are amplitudes of such modes. In both cases, the eikonal action arises as a limit of a local effective action for the reparametrization modes. As it stands, the actions we derived in cases with non-maximal chaos are non-local. It would be very desirable to construct a local effective field theory that generalizes these actions to the case of non-maximal chaos, with degrees of freedom that are suitably generalized reparametrization with $X^\pm$ the amplitude of scramblon modes. One noteworthy point is the fact that maximally chaotic systems typically have additional symmetries. Even the saddle point of the large $q$ SYK model still has an $SL(2,\mathbb{R})$ symmetry, as we discussed, and this can be used to generate field configurations. However, in the large $q$ SYK chain such a symmetry seems genuinely absent, which makes this case particularly interesting. Relatedly, we note that the vertex factors in the large $q$ chain are rather complicated and not generated by reparametrizations (at least not in any obvious way). The $SL(2,\mathbb{R})$ is interpreted geometrically as a ``fake thermal circle'' in \cite{StanfordTalk}. From that perspective it is very interesting that non-maximal chaos in the SYK chain originates from two sources: a ``fake thermal circle'' and an inherently stringy non-geometric component.

These vertex factors in turn gave rise to very interesting new effects in the OTOC in \eqref{eq:SeikFinal2}, which we tentatively interpreted as coming from multi-string exchange. Away from the maximal chaos these effects are not suppressed. It would be very interesting to understand them in field theories with a known holographic string dual.

\section*{Acknowledgments}

We thank A.\ Streicher for initial collaboration on this project. We are also grateful to M.\ Blake, A.\ Kitaev, H.\ Lin, W.\ Reeves, M.\ Rozali, D.\ Stanford, Y.\ Zhao for useful discussions. We thank P.\ Gao and H.\ Liu for coordinating the {arXiv} submission of their paper~\cite{Gao:2023wun} with us. The research of CC was supported by the Perimeter Institute for Theoretical Physics. Research at Perimeter Institute is supported in part by the Government of Canada through the Department of Innovation, Science and Economic Development and by the Province of Ontario through the Ministry of Colleges and Universities. FH acknowledges support provided by the DOE grant DE-SC0009988 and by the UKRI Frontier Research Grant EP/X030334/1.

\appendix

\section{$SL(2,\mathbb{R})$ symmetry of large $q$ SYK}
\label{sec:SL2}

In this appendix we expand on the $SL(2,\mathbb{R})$ symmetry, which is preserved by the saddle point solution of the large $q$ SYK model. For ease of notation we focus on the Euclidean expression, which we write as
\beq
e^{g(s_1,s_2)}={1\ov \mathcal J^2}{F'(s_1) G'(s_2)\ov (F(s_1)-G(s_2))^2} \,.
\eeq
These solutions are invariant under 
\beq
 SL(2,\mathbb{R})_\text{diag}: \qquad  F(s) \; \longrightarrow \; \frac{a\, F(s) +b}{c\, F(s) + d}\,,\qquad G(s) \; \longrightarrow \; \frac{a \,G(s) +b}{c \,G(s) + d} \,,\qquad ad-bc=1\,.
\eeq
The infinitesimal symmetry generators are $L_n^\text{(diag)} = L_n^{(1)} + L_n^{(2)}$, where the two terms act separately on the two time coordinates:
\beq
\begin{split}
  L_n^{(1)} \, e^{\Delta g(s_1,s_2)} &= -i e^{-in \left( v s_1 + \frac{1-v}{2} \, \pi \right)} \left( \frac{1}{v}\,\partial_{s_1} -i\Delta n \right) e^{\Delta g(s_1,s_2)} \,,\\
   L_n^{(2)} \, e^{\Delta g(s_1,s_2)} &= -i e^{-in \left( v s_2 - \frac{1-v}{2} \, \pi \right)} \left(\frac{1}{v}\, \partial_{s_2} -i\Delta n \right) e^{\Delta g(s_1,s_2)} \,.
\end{split}
\eeq
By introducing twisted coordinates $\tilde{s}_1 = v s_1 + \frac{1-v}{2} \, \pi$ and $\tilde{s}_2 = v s_2 - \frac{1-v}{2}\, \pi$, these generators simplify and take a more familiar form:
\beq
  \tilde{L}_n^{(a)} = - i e^{-in \, \tilde{s}_a} \left( \partial_{\tilde{s}_a} - i\Delta n \right) \qquad (a=1,2)\,.
\eeq
That is, the $SL(2,\mathbb{R})_\text{diag}$ symmetry can be uniformized by twisting the circle coordinates. This trick has appeared before (e.g., \cite{Streicher:2019wek}) and is useful in calculations.

\section{Soft modes and eikonal action in Brownian SYK}\label{app:Brownian}

In this appendix we review the derivation of the eikonal action for the Brownian SYK model, following the discussion in \cite{Stanford:2021bhl}.
We start with Brownian SYK linearized around the saddle point on the 4-fold contour. We are looking for soft modes of $K-S$, as in (B.9)-(B.11) of \cite{Stanford:2021bhl}, which we adapt to our purposes:
{\small
\es{KmS}{
(K-S)\star \vec{\sig}&=\mu\, \vec{\sig}\\
\begin{pmatrix}
-\sig_x(t_1)+{(q-1)\lam_L\ov q-2}\int_{-T/2}^{t_1}dt_2\ e^{-{\lam_L\ov q-2} t_{12}}\le(\sig_y(t_2)-\sig_z(t_2)\ri)\\
\sig_y(t_1)-{(q-1)\lam_L\ov q-2}\le[\int_{-T/2}^{t_1}dt_2\ e^{-{\lam_L\ov q-2} t_{12}}\le(\sig_y(t_2)-\sig_z(t_2)\ri)+\int_{t_1}^{T/2}dt_2\ e^{-{\lam_L\ov q-2} t_{21}}\le(\sig_y(t_2)-\sig_x(t_2)\ri)\ri]\\
-\sig_z(t_1)-{(q-1)\lam_L\ov q-2}\int_{t_1}^{T/2}dt_2\ e^{-{\lam_L\ov q-2} t_{21}}\le(\sig_y(t_2)-\sig_x(t_2)\ri)
\end{pmatrix}&=\mu\begin{pmatrix}
\sig_x(t_1)\\
\sig_y(t_1)\\
\sig_z(t_1)
\end{pmatrix}
}
}\normalsize
These equations can be solved exactly, and the solution takes the schematic form:
\es{Eigen}{
\begin{pmatrix}
\sig_x(t)\\
\sig_y(t)\\
\sig_z(t)
\end{pmatrix}&=
\begin{pmatrix}
A_x \sinh(\Lam(t+T/2))\\
{1+\mu\ov 1-\mu}\le[A_x \sinh(\Lam(t+T/2))+A_z \sinh(\Lam(T/2-t))\ri]\\
A_z \sinh(\Lam(T/2-t))
\end{pmatrix}\,,
}
where
\es{LammuRel}{
\Lam(\mu)&=\lam_L\sqrt{\le({1+{\mu\ov q-2}\ov 1-\mu}\ri)^2-\le({q-1\ov q-2}\ri)^2\,\le({2\mu \ov 1-\mu^2}\ri)^2}\\
&=\lam_L\le(1+{q-1\ov q-2}\,\mu+O(\mu^2)\ri)\,.
} 
(We will not need the expression for $A_z/A_x$.) The integral equation \eqref{KmS} determines its own boundary conditions, from the first component $\sig_x(-T/2)$ and from the third $\sig_z(T/2)=0$, which leads to the quantization of 
 $\mu$
through the equation:
\es{quantCond}{
\Lam\coth(\Lam T)=\lam_L\,{1+{\mu\ov q-2}\ov 1-\mu}\,.
}
For the derivation above, refer to Appendix B of \cite{Stanford:2021bhl}. From \eqref{quantCond} we can find two near-zero modes:
\es{smallmu}{
\mu_{S,A}=\pm{q-2\ov q-1} e^{-\lam_L T}\,.
}

Now we make an interesting observation. In the large $T$ limit, for $t=O(1)$ the eigenfunctions \eqref{Eigen} corresponding to these eigenvalues become:
\es{SmallEigenVec}{
\vec{\sig}_{S,A}&=\vec{\Sig}_{S,A}+O(e^{-\lam_L T})\,,\\
\vec{\Sig}_{S,A}&=
\begin{pmatrix}
e^{\lam_L t}\\
e^{\lam_L t}\pm e^{-\lam_L t}\\
\pm e^{-\lam_L t}
\end{pmatrix}\propto\le[\de_{x^{+}} \vec{\sig}(x^{+})\pm \de_{x^{-}} \vec{\sig}(x^{-})\ \ri]\,,
}
where $\vec{\sig}(x^{\pm})$ is the one shock saddle point (given in (B.19)-(B.20)), which we reproduce here:
\es{sigpm}{
\vec{\sig}(x^{+})=-{\lam_L\ov 2(q-2)}\begin{pmatrix}
\le({1\ov 1+e^{\lam_L (t-T/2)}\,x^+}\ri)^{q-1\ov q-2}\\
\le({1\ov 1+e^{\lam_L (t-T/2)}\,x^+}\ri)^{q-1\ov q-2}\\
0
\end{pmatrix}\,,\quad \vec{\sig}(x^{-})=-{\lam_L\ov 2(q-2)}\begin{pmatrix}
0\\
\le({1\ov 1+e^{-\lam_L (t+T/2)t}\,x^-}\ri)^{q-1\ov q-2}\\
\le({1\ov 1+e^{-\lam_L (t+T/2)}\,x^-}\ri)^{q-1\ov q-2}
\end{pmatrix}
}
$\vec{\Sig}_{S,A}$ are zero eigenvectors of $(K-S)_\infty$.

If we are interested in the small eigenvalue \eqref{smallmu} we do not need to go through the complicated procedure of solving the problem exactly. Instead, we can regard $(K-S)_T$ as integral operators on the infinite line, but with a kernel localized to $(-T/2,T/2)$, and apply first order (degenerate) perturbation theory. Then we get
\es{FirstOrder}{
\lam_{S,A}={\bra{\vec{\Sig}_{S,A}}(K-S)_T\ket{\vec{\Sig}_{S,A}}\ov \braket{\vec{\Sig}_{S,A}}{\vec{\Sig}_{S,A}}}\,,
}
where we have chose the basis in the two-dimensional degenerate eigenspace of $(K-S)_\infty$ such that $\bra{\vec{\Sig}_{S}}(K-S)_T\ket{\vec{\Sig}_{A}}=0$. Performing the computation of \eqref{FirstOrder} gives perfect agreement with \eqref{smallmu}.

We can now write the effective action: 
\es{action}{
I&=-N{q-2\ov (q-1)\lam_L} \vec{\sig} \star (K-S)_T\star \vec{\sig}\\
&=-{N\ov 2(q-2)^2}\, e^{-\lam_L T}\, x^+x^-+\text{(higher order)}\,.
}
We use the same ideas in regular Hamiltonian SYK.

\section{Analysis of the large-$q$ OTOC} \label{ruelle}

The result \eqref{eq:FotocRes} has appeared in the literature before (in particular \cite{Gu:2021xaj}), but as far as we know it has not been analyzed in detail.\footnote{ See however ref.\ \cite{Susskind:2022bia}, where the related quantity operator size was discussed using the epidemic model approximation. It dispalys features similar to our OTOC.}  Let us therefore discuss some of its features. First, we note that the expansion \eqref{eq:FotocRes22} is somewhat misleading: if $\Delta$ was ${\cal O}(1)$ (as in the typical analysis of the Schwarzian, see section \ref{sec:schwarzian} or \cite{Maldacena:2016upp}), then one would conclude that scrambling occurs with exponential time dependence and the characteristic time scale for the exponentially growing term to become significant would be identified as 
\begin{equation}
t_\text{scr} \sim \frac{1}{v} \log \left(\frac{N}{q^2}\right)\,.
\end{equation}
This time scale is still relevant: it corresponds to the time over which the OTOC decays exponentially from its initial value 1 with a Lyapunov exponent $\lambda_L = \frac{2\pi v}{\beta}$. What is unusual compared to examples such as the Schwarzian, is that this is not the time scale it takes for the OTOC to deviate from $1$ by an ${\cal O}(1)$ amount: after a scrambling time $t_\text{scr}$ the OTOC has merely decayed by a small amount to $1-\frac{1}{q^2}$.

We refer to the regime after the scrambling time as the Ruelle regime, where the OTOC approaches zero according to a power law. For very small values of $\Delta$ (such as $\Delta_\psi$), after a time $t_\text{scr}$ when exponential scrambling ends the OTOC is still far from zero: the Ruelle regime is enhanced and lasts for a very long time. For small $\Delta$, the hypergeometric function is well approximated by the first terms in the large $z$ expansion:
\begin{equation}
\label{eq:UexpandDelta}
\begin{split}
z^{-2\Delta} \, U\big(2\Delta,1, z^{-1}\big) 
&\approx z^{-2\Delta} \; \left[ \frac{\log z - \psi(2\Delta)-2\gamma_E}{\Gamma(2\Delta)} + {\cal O}\left( \frac{1}{z} \right) \right] \\
&\approx 1 - \Delta^2 \left[ 2 \log^2 z - 4 \gamma_E \, \log z + \pi^2 + 2 \gamma_E^2 + {\cal O}\left(\frac{1}{z}\right) \right]  +  {\cal O}(\Delta^3) 
\end{split}
\end{equation}
where $\psi(x) \equiv \frac{\Gamma'(x)}{\Gamma(x)}$, $\gamma_E$ is the Euler-gamma, and we expanded additionally in small $\Delta$ in the second line.\footnote{ Note that this expansion is convergent. Including more terms as indicated in \eqref{eq:UexpandDelta} yields a successively better approximation.} For small $\Delta$, the first line of \eqref{eq:UexpandDelta} provides an excellent approximation for almost all times except the very early scrambling regime. The decay in this Ruelle regime is polynomial in $vT$. The characteristic time according to the above approximation is found by solving the condition $2\Delta^2 \log^2 z  + \ldots \approx {\cal O}(1)$:
\begin{equation}
\label{eq:largeqScr}
 t_\text{Ruelle} \sim t_{scr} + \frac{c_1}{v\Delta} + {\cal O}(1) 
\end{equation}
with $c_1 \sim {\cal O}(1)$. While the dependence of scrambling time is logarithmic on $N$ (and $q$), the other term is linear in $q$ for $\Delta= {\cal O}(1/q)$. 

\begin{figure}[!h]
       \begin{center}
       \includegraphics[width=.7\linewidth]{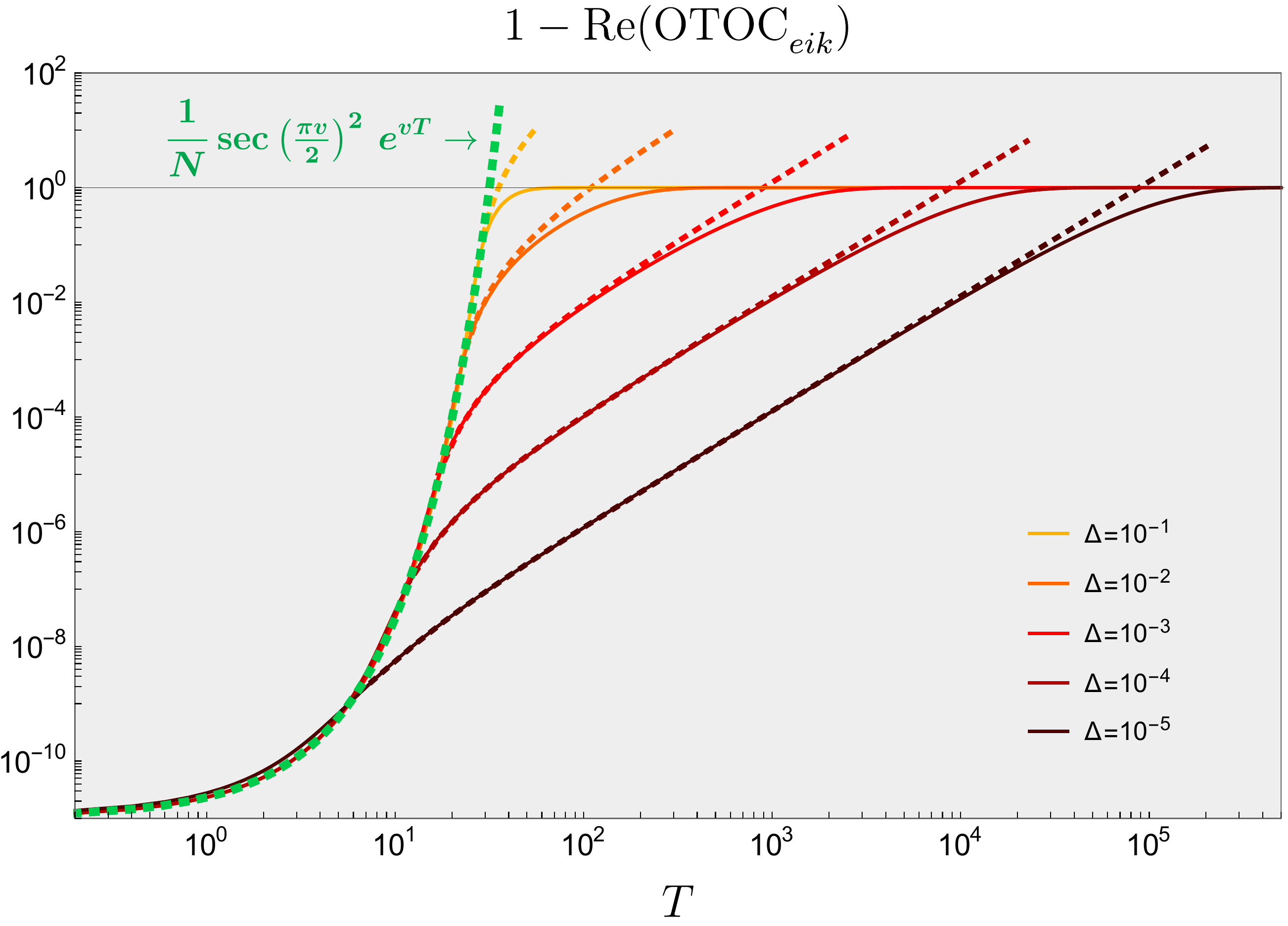}
       \end{center}\vspace{-.2cm}
     \caption{Doubly logarithmic plot of the long-time OTOC \eqref{eq:FotocRes} of the large $q$ SYK model for $N=10^{12}$, $v=0.8$, and some values of $\Delta$. 
  The dotted green line shows the approximately exponential time dependence $\sim e^{vT}$, which is applicable for all values of $\Delta$ at early times.
     Other dotted lines approximate the subsequent `Ruelle' regime up to ${\cal O}(\Delta^2)$, as shown in the second line of \eqref{eq:UexpandDelta}. The approximation in the first line of \eqref{eq:UexpandDelta} is not shown as it would be indistinguishable from the solid lines in the Ruelle regime.
     One can verify from the plot that `scrambling' ends after a time of order $t_\text{scr} \sim \frac{1}{v} \log(N/q^2)$ and
     the time for the OTOC to deviate from 1 by an ${\cal O}(1)$ amount scales with $\Delta^{-1}$, c.f., \eqref{eq:largeqScr}.}
     \label{fig:Uplot}
\end{figure}

We illustrate all these observations in figure \ref{fig:Uplot}: it shows the characteristic time scale for the Ruelle regime, $t_\text{Ruelle} \sim \Delta^{-1} $, which determines the decay after the scrambling regime (green line), which lasts for a time of order $t_\text{scr} \sim \log(N/q^2)$. Note that for the purpose of this figure we took $N \gg q^2 \gg \log (N) \gg 1$. To be precise, this is not exactly the limit we are analyzing in the rest of this paper, as we should take the strict large $N$ limit first. The plot nevertheless illustrates some of the features we discuss above.

We take all this as evidence that the OTOC built out of single-fermion operators behaves in slightly unusual way.
To remedy this, we will in the following consider OTOCs built out of clusters of many $s_1$ and $s_2$ fermions, respectively, such that the operators in the four-point function effectively have dimension $\Delta_{s_i} \equiv s_i \Delta_\psi \equiv \frac{s_i}{q}$.

\section{Leading connected OTOC from ladder diagrams (comparison with \cite{Gu:2018jsv})} \label{GKcompare}

The asymptotic form of the OTOC for the SYK chain was computed in \cite{Gu:2018jsv}, based on general properties of the OTOC in theories dominated by ladder diagrams. We briefly review their derivation in order to contrast with ours and to confirm that the results match. For details on the diagrammatic arguments we refer to the references.

In the diagrammatic computation of the OTOC in SYK (i.e., solving the recursive Schwinger-Dyson equation for the four-point function), a crucial role is played by the ladder kernel \cite{Maldacena:2016upp}. In the SYK chain, its retarded ladder kernel at $q\rightarrow \infty$ takes the following form:
\begin{equation}
\label{eq:KRAdef}
\begin{split}
K^{R}(t_1,t_2,t_3,t_4;p) 
  & = \;\frac{h(h-1)}{2}\left[ \Theta(t_{13}) \Theta(t_{24}) \,\frac{v^2}{2\cosh^2\left( \frac{vt_{34}}{2} + \frac{i\pi v}{2} \right)} \right]
  \\&= \frac{h(h-1)}{2}\, K^{R}(t_1,t_2,t_3,t_4;0) 
 \end{split}
\end{equation}
where the ladder is determined from the Bethe-Salpeter equation $\mathcal F(p)=\mathcal F_0(p)+K(p)\mathcal F(p)$ for each momentum sector $p$, 
and we took the large $q$ limit. The advanced ladder kernel $K^A$ takes a similar form.

Out of the ladder kernel, we build the following operator:
\begin{equation}
\begin{split}
 K_\alpha^{R}(t,t';p) = \int ds \, K^{R}\left( s + \frac{t}{2} , \, s- \frac{t}{2} , \,\frac{t'}{2} ,\,- \frac{t'}{2} ;p\right) \, e^{\alpha s}\,.
\end{split}
\end{equation}
This kernel determines some important quantities: the Schwinger-Dyson equation implies that ${\cal F}^\text{otoc}\big(\frac{t'}{2},-\frac{t'}{2},t_3,t_4;p\big)$ is an eigenfunction of the operator $K_\alpha^{R}$ with eigenvalue 1. The Lyapunov exponent is given by the value of $-\alpha$ for which this is satisfied. The retarded/advanced eigenfunctions of this kernel and their respective eigenvalues for the chain work out to be:\footnote{The eigenvalue $k_R(\alpha;p)$ can be obtained from the eigenvalue in SYK after taking $q\rightarrow \infty$ and $\alpha \rightarrow \alpha/v$.}
\beq
\Upsilon^{R}_\alpha(t)=\Upsilon^{A}_\alpha(t)=\left[2\cosh\left ( {v t\ov 2}+{i\pi v \ov 2} \right)\right]^{\al\ov v},\qquad k_{R}(\al;p) =\frac{h(h-1)}{2}  \,  {2v^2 \ov \al(\al-v)} \,.
\eeq
Indeed, the Lyapunov exponent $\kappa(p)$ determined this way has the expected value:
\begin{equation}
  k_R\big(\alpha=-\kappa;p\big) \stackrel{!}{=} 1 \qquad \Rightarrow \qquad  \kappa(p) = v(h-1) \,.
\end{equation} 

These considerations determine the time dependence of the OTOC: being an eigenfunction of the retarded ladder kernel $K^R$ with eigenvalue $1$, its momentum modes can be written as\footnote{The phase $e^{-\kappa i \pi}$ has the opposite sign in the exponent compared to \cite{Gu:2018jsv} (see also \cite{Kitaev:2017awl}). This is because we compute an OTOC in the configuration $0 = \frac{\hat{t}_1 + \hat{t}_2}{2}  \ll \frac{\hat{t}_3 +\hat{t}_4 }{2} = T$ whereas \cite{Gu:2018jsv} consider $\hat{t}_1 + \hat{t}_2  \gg \hat{t}_3 +\hat{t}_4 $ (but with the same ordering of the insertion times along the contour). Effectively, our large $T$ would have the opposite sign in their conventions. This has the effect of changing the phase of the OTOC, as one can easily verify in examples such as \eqref{eq:FOTOClargeq}.}
\begin{equation}
\label{eq:FotocGuKitaev}
 \frac{1}{N}\,{\cal F}^\text{otoc}(\hat{t}_1,\hat{t}_2,\hat{t}_3,\hat{t}_4;p) \approx -\frac{e^{\kappa(-i\pi + \hat{t}_3+\hat{t}_4-\hat{t}_1-\hat{t}_2)/2}}{C(p)} \, \Upsilon^R(\hat{t}_{12};p)\Upsilon^A(\hat{t}_{34};p) \,,
\end{equation} 
where $\Upsilon^{R/A}(t;p) \equiv \Upsilon^{R/A}_{\alpha=-\kappa(p)}(t)$. To determine the missing momentum-dependent prefactor $C(p)$, ref.\ \cite{Gu:2018jsv} derives the {\it ladder identity}:\footnote{ The ladder identity differs by a factor of ${1\ov 4}$ from \cite{Gu:2018jsv}: our pre-factor is $\frac{N}{2}$ instead of $2N$. The extra factor of $\frac{1}{4}$ is due to our definition of ${\cal F}$ as the leading connected four-point function {\it after} normalizing by two-point functions, see \eqref{eq:OTOCconfigDef}. 
}
\begin{equation}
\label{eq:ladderid}
\begin{split}
 C(p) &= \frac{N}{2} \cdot\cos \left( \frac{\pi \kappa}{2} \right) \cdot t_B \, (\Upsilon^A,\Upsilon^R)
\end{split}
\end{equation} 
where 
\begin{equation}
 t_B = \frac{d}{d\alpha} k_R(\alpha;p) \Big{|}_{\alpha=-\kappa} 
 = \frac{(2h-1)}{vh(h-1)}
 \end{equation}
is the branching time, and the inner product of vertex functions is
\begin{equation}
\begin{split}
  (\Upsilon^A,\Upsilon^R) 
  &\equiv \int dt\, \Upsilon^A(t;p) \frac{ {{h(h-1)\ov 2}} v^2}{2\cosh^2\! \left( \frac{vt}{2} +\frac{i\pi v}{2} \right)} \Upsilon^R(t;p) \\
  &={{h(h-1)\ov 2}}  \frac{2v}{(2h-1)4^{h-1}\sqrt{\pi}} \, \cos \left(\frac{\pi \kappa}{v}\right) \Gamma
   \left(\frac{\kappa}{v}+1\right) \Gamma \left(\frac{1}{2}-\frac{\kappa}{v}\right)
  \end{split}
\end{equation} 
Finally, after Fourier transforming to $x$-space the leading term in the OTOC takes the following form:
\beq
\begin{split}
\frac{1}{N}{\cal F}^\text{otoc}(\hat{t}_1,\hat{t}_2,\hat{t}_3,\hat{t}_4;x) &= -\int_{-\infty}^{\infty}{dp\ov 2\pi} \,{e^{ipx+\kappa(-i\pi+\hat{t}_3+\hat{t}_4-\hat{t}_1-\hat{t}_2)/2}\ov C(p)}\,\Upsilon^R(\hat{t}_{12};p)\Upsilon^A(\hat{t}_{34};p)  \\
 &= -\frac{2\sqrt{\pi}}{N} \int_{-\infty}^{\infty}{dp\ov 2\pi} \, {e^{ipx+\kappa \left(- \frac{i\pi}{2} +T\right)}\ov \cos\left({\pi \kappa\ov 2}\right) \cos \left( \frac{\pi \kappa}{v} \right) \Gamma\left( \frac{\kappa}{v}+1 \right) \Gamma\left( \frac{1}{2} - \frac{\kappa}{v} \right)}\left[ {1 \ov \cos \left( \frac{\pi v}{2} \right) }\right]^{\frac{2\kappa}{v}}
\end{split}
\eeq

\bibliographystyle{JHEP}
\bibliography{refs.bib}

\end{document}